%% file: FlapMethods.tex
\documentclass[3p]{elsarticle}

\usepackage{graphicx, amsmath, amssymb, amsfonts, mathtools, mathrsfs, color}
\usepackage{comment, enumerate, tabularx, multirow}
\usepackage{natbib, hyperref, url}
\usepackage{algorithm, algpseudocode, pifont, longtable}

\newcommand{\vsp}[1]{\vspace{#1 pc} \noindent}

\newcommand{\newa}{}		
\newcommand{\newb}{}		
\newcommand{\td}[2]{\frac{d #1 }{d #2}}

\newcommand{\pd}[2]{ \frac{ \partial #1}{ \partial #2 } }
\newcommand{\ppd}[2]{ \frac{ \partial^2 #1}{ {\partial #2}^2 } }
\newcommand{\BigO}[1]{ \mathcal{O} \left(#1\right) }
\newcommand{\bvec}[1]{\ensuremath{\boldsymbol{#1}}}
\newcommand{\grad}{\nabla} 
\newcommand{\abs}[1]{\left| #1 \right|}
\newcommand{\tavg}[1]{\langle #1 \rangle}
\newcommand{\norm}[1]{ \left\| #1 \right\| }
\newcommand{\ip}[1]{ \langle #1 \rangle }

\newcommand{\freq} {f}
\newcommand{\chord}{c}
\newcommand{\Udim}{U_{\infty}}
\newcommand{\thick}{ b }
\newcommand{\amp}{ A }
\newcommand{\width}{ w }
\newcommand{\Ia}{I}
\newcommand{\UFS}{ U }

\newcommand{\Mrat}{ R }
\newcommand{\Stiff}{ S }
\newcommand{\St}{\text{St}}
\newcommand{\pinf}{ p_{\infty} }
\newcommand{\loadDim}{{q}}
\newcommand{\load}{\tilde{q}}
\newcommand{\loadx}{Q}
\newcommand{\loadsing}{\loadx_s}
\newcommand{\loadreg}{ \loadx_{r} }
\newcommand{\singterm}{ \sqrt{\frac{1-x}{1+x}} }
\newcommand{\sinsum}{ \sum_{k=1}^{\infty} \co_k \sin {k \theta} }
\newcommand{\bu}{\bvec{u}}
\newcommand{\ejt}{e^{2 \pi j t}}
\newcommand{\etah}{\hat{\eta}}
\newcommand{\etalin}{ \lambda }
\newcommand{\etaLE}{\eta_{ _\text{LE} }}
\newcommand{\vn}{ \hat{V} }
\newcommand{\theofun}{C}
\newcommand{\compot}{F}
\newcommand{\paccel}{\phi}
\newcommand{\wing}{\mathcal{W}}
\newcommand{\ucirc}{\mathcal{C}}
\newcommand{\dtOpp} {\left( \pd{}{t} + \UFS \pd{}{x} \right)}
\newcommand{\Dx}{ \mathcal{D} }
\newcommand{\Dinv}{\Dx^{-1} }
\newcommand{\Popp}{ \mathscr{P} }
\newcommand{\Pinv}{ \Popp^{-1} }
\newcommand{\sump}[2]{\sum_{#1}^{#2} {}^{'} }
\newcommand{\Lopp}{\mathscr{L}}
\newcommand{\trialf}{\tilde{\eta}}
\newcommand{\forxc}{\text{for } x \in [-1,1]}
\newcommand{\forxo}{\text{for } x \in (-1,1)}
\newcommand{\Tdim}{T}
\newcommand{\Pdim}{P}
\newcommand{\Tnd}{\tilde{T}}
\newcommand{\Pnd}{\tilde{P}}

\newcommand{\CT}{ {C}_\text{T} }
\newcommand{\CP}{ {C}_\text{P} }
\newcommand{\etaref}{\eta_{\text{ref}}}
\newcommand{\co}{a}		
\newcommand{\pco}{b}		
\newcommand{\Pco}{B}		
\newcommand{\ppco}{b'}		
\newcommand{\chico}{c}		
\newcommand{\Chico}{C}		
\newcommand{\alco}{d}		
\newcommand{\Alco}{D}		
\newcommand{\MyComment}{$\triangleright$ Comment: }

\journal{Journal of Computational Physics}
\begin{document}
\title{A fast Chebyshev method for simulating flexible-wing propulsion}
\author{M. Nicholas J. Moore}

\begin{abstract}
We develop a highly efficient numerical method to simulate small-amplitude flapping propulsion by a flexible wing in a nearly inviscid fluid. We allow the wing's elastic modulus and mass density to vary arbitrarily, with an eye towards optimizing these distributions for propulsive performance. The method to determine the wing kinematics is based on Chebyshev collocation of the 1D beam equation as coupled to the surrounding 2D fluid flow. Through small-amplitude analysis of the Euler equations (with trailing-edge vortex shedding), the complete hydrodynamics can be represented by a nonlocal operator that acts on the 1D wing kinematics. A class of semi-analytical solutions permits fast evaluation of this operator with $\BigO{N \log N}$ operations, where $N$ is the number of collocation points on the wing. This is in contrast to the minimum $\BigO{N^2}$ cost of a direct 2D fluid solver. The coupled wing-fluid problem is thus recast as a PDE with nonlocal operator, which we solve using a preconditioned iterative method. These techniques yield a solver of near-optimal complexity, $\BigO{N \log N}$, allowing one to rapidly search the infinite-dimensional parameter space of all possible material distributions and even perform optimization over this space.
\end{abstract}
\maketitle

\section{Introduction}

	A variety of flying and swimming animals propel themselves by flapping wings or fins in a fluid. Inspired by their agility, stealth, and efficiency \cite{Anderson1998, Tangorra2007, Shang2009}, researchers have been working to integrate similar principles into new technologies, such as autonomous underwater vehicles, ornithopters, and micro-air vehicles \cite{DeLaurier1999, DeLaurier2003, Liu2010, Curet2011, Moored2011a, Moored2011b, Whitney2012, Ristroph2014}

	A distinguishing feature of natural locomotion is the use of flexible wings or fins. These appendages deform significantly when actuated, which can provide a number of performance benefits. In addition, natural wings and fins often exhibit complex material compositions, with elastic properties that can be highly nonuniform and even anisotropic \cite{Khan2009, Wood2011, Zhu2013, Lucas2015}. Understanding how biology exploits material heterogeneity could significantly advance the design of artificial devices. However, such an understanding not only hinges on a highly complex fluid-structure interaction, but also involves the vast parameter space of all possible material distributions.

	Recently, great theoretical, experimental, and computational efforts have been directed towards advancing our understanding of flexible-wing propulsion \cite{Gursul2007, Alben2008, Alben2009vs, Spagnolie2010, Thiria2010, Shelley2011, Dai2012, Moored2014, Tian2014, Quinn2014}. These studies have demonstrated that flexibility can drastically improve propulsive performance, especially when a wing or fin is driven near resonance \cite{Mich2009, Masoud2010, Dewey2013, Moore2014, Paraz2016}. Several studies even performed optimization to uncover wing properties and/or actuation strategies that deliver peak performance \cite{Tuncer2005, Alben2008, Moored2014, Quinn2015, Paraz2016}. All of the studies mentioned above, though, considered only the simplest material distributions, such as uniform flexibility \cite{Gursul2007, Alben2008, Alben2009vs, Mich2009, Masoud2010, Dewey2013, Moored2014, Paraz2016} or flexibility that is localized through a torsional joint \cite{Spagnolie2010, Moore2014}.

	Research on heterogeneous wings has appeared more recently \cite{Wood2011, Nakata2012, Nakata2012Proc, Zhu2013, Curet2014, Lucas2015, Liu2016, JFS2016}. The majority of these studies focus on insect flight and, more specifically, simulating the coupled aeroelastic dynamics of compositionally complex wings \cite{Nakata2012, Nakata2012Proc, Zhu2013, JFS2016}. Since they employ direct numerical simulation (DNS) of the coupled Navier-Stokes/elastic-body partial differential equations (PDEs), the computational cost of a single simulation can be quite high. As such, the simulations can only examine a handful of different material distributions, typically chosen to mimic real insect wings. These simulations demonstrate definite performance advantages of insect-like distributions over stiff wings \cite{Nakata2012Proc, Zhu2013}. However, it remains to be seen whether these distributions are optimal in any sense, as the infinite-dimensional space of arbitrary material distribution remains largely unexplored by these methods.

	In contrast, Moore (2015) constructed an asymptotic-based method for flexible-wing propulsion  with the computational efficiency needed to rapidly search material-distribution space \cite{Moore2015}. The high efficiency even renders numerical optimization feasible. For the case of a wing in forward flight (i.e.~heaved vertically at the leading edge and steadily translating in the horizontal direction), it was shown that concentrating flexibility near the driving point can enhance thrust production significantly. A torsional spring can be used to focus all of the flexibility at one point, and this arrangement was found to globally optimize thrust \cite{Moore2015}. Intriguingly, this finding is consistent with the architecture of insect wings, in which the elastic wing-body joint acts as a torsional spring positioned near the wing's leading edge, and the majority of elastic deformations occur along that axis \cite{Ennos1988a, Wood2011}.

	At the heart of Moore's optimization procedure lies a highly efficient PDE solver to determine the wing kinematics resulting from a given actuation strategy and wing composition \cite{Moore2015}. The solver achieves its efficiency by using {\em small-amplitude asymptotics} to describe the flow field. Vortex shedding is taken into account by enforcing a Kutta condition, and the resulting flow couples to the wing's bending motions through Euler-Bernoulli beam theory. While this PDE solver was used in the optimization procedure of Moore (2015) \cite{Moore2015} and subsequent work on flapping membranes \cite{Alon2017}, the underlying analysis was not discussed in any detail. 
	
	The purpose of this paper is to detail the mathematical features that give rise to the method's speed and accuracy. In particular, we show that the collective influence of the hydrodynamics collapses to a {\em nonlocal} operator that acts on the 1D wing kinematics. We introduce a method to rapidly evaluate this operator with $\BigO{N \log N}$ operations, where $N$ is the number of nodes on the wing. Recasting the PDE system in terms of the nonlocal operator not only reduces the dimensionality but also reveals an interesting Riemann-Hilbert structure. We solve the resulting boundary-value problem (BVP) with a preconditioned iterative method, where the preconditioning is performed with {\em continuous} operators {\newb (similar to an integral reformulation \cite{Greengard1991})}. This approach allows analytical removal of a flow singularity that exists at the wing's leading edge. We present new analysis, based on careful examination of higher-order singularities, to quantify the accuracy of the method, and we benchmarks results against newly derived asymptotic solutions. We also discuss a few additional applications, such as different actuation strategies and performance metrics, to supplement the results of Moore (2015) \cite{Moore2015}.
	
	The main difference between the method discussed here and other inviscid, vortex-shedding methods \cite{Cottet2000, Katz2001, Jones2005, Mich2009, MichMethod2009, Alben2009vs, Alben2010vsa, Alben2010vsb} is that, {\newa within the small-amplitude regime}, there is no need to track the trailing vortex sheet. Since its length grows indefinitely with time, many methods that do track the sheet require an ever-increasing amount of computational effort. This effect can severely slow simulations, particularly if pair-wise vortex interactions are handled directly with $\BigO{N^2}$ operations. 
{\newa In the case of large-amplitude flapping, there are few alternatives, as nonlinear effects in the wake (e.g.~vortex-sheet rollup) can influence propulsor dynamics. However, in many applications, such as bird-flight based ornithopters \cite{DeLaurier1999, DeLaurier2003} or Carangiform/Thunniform swimming \cite{Alben2008}, the amplitude of flapping is small compared to a propulsor length-scale, which allows the governing Euler equations to be linearized. Following the original small-amplitude formulation of Wu (1961) \cite{Wu1961}, our method features the {\em pressure field} as the primary unknown. Since pressure is continuous everywhere in the fluid domain, in particular across the vortex sheet, there is no need to track this sheet.
Building on Wu's work, which either considered {\em prescribed} kinematics \cite{Wu1961} or determined certain kinematics that minimize elastic recoil (but without determination of the underlying inter-muscular forces) \cite{Wu1971}, we merge the small-amplitude theory with well-developed BVP solvers (see for example \cite{Greengard1991, GreenRokh1991, Shen1994, Shen1995, Olver2013}) to determine the emergent deformations of a flexible propulsor. Of particular importance is the representation of the hydrodynamics as a nonlocal operator in the BVP governing deformations.
}

	In closely related work, Alben (2008) devised a method to simulate flexible-appendage propulsion  in the same small-amplitude regime considered here \cite{Alben2008, Alben2009sa, Alben2010sa, Alben2012saa, Alben2012sab}. That method, though, describes the flow in terms of the vorticity field, more like the studies mentioned above. The small-amplitude linearization allows the wake circulation to be precomputed, thus eliminating the need to track the trailing vortex sheet and providing a computational speed-up. The unknowns that remain are the bound vorticity and wing deformations. These two quantities are linked through pressure; knowledge of the vorticity field allows computation of the pressure, which then acts as the load that gives rise to wing deformations and, ultimately, thrust generation. When discretized, these relationships produce a dense linear system which Alben solved iteratively with $\BigO{N^2}$ operations.
		
	In contrast, our method describes the flow in terms of the pressure field {\em directly} without going through vorticity as an intermediate variable. This approach leads to gains in {\em computational efficiency}, as we can iteratively solve for the emergent wings kinematics with $\BigO{N \log N}$ operations. We argue that this approach also offers {\em conceptual simplicity}, as it is the pressure that directly links hydrodynamic forces and wing deformations. In many applications, there is no need to compute the vorticity field at all. Once the kinematics and pressure distributions are known, the thrust generated by the wing and other performance metrics follow as simple calculations. Thus, our method enables the computation of certain high-Reynolds-number fluid-structure interactions with computational efficiency similar to that seen in fast Stokes solvers based on singularity methods \cite{Tornberg2006, Tornberg2008, Keaveny2011, Jiang2012, Gimbutas2015}. Since their advent, these solvers have enabled rapid advances in low-Reynolds-number applications \cite{Fauci2006, Kanevsky2010, Spagnolie2015, Mitchell2016, Elfring2016}, for instance shape optimization of micro-swimmers \cite{Keaveny2013}. Similar advances could and should be expected for analogous high-Reynolds-number problems.
	
	The paper is organized as follows. In Section  \ref{FormulationSection} we describe the physical setup and introduce the small-amplitude linearization of the fluid equations. In Section  \ref{FluidSection} we derive semi-analytical solutions for the flow field based on the assumption of {\em prescribed} wing kinematics. In Section \ref{BendingSection}, we couple this flow to the initially unknown kinematics of a {\em flexible} wing, and we describe a Chebyshev-based method to solve for the resulting deformations. In this section, we introduce the {\em nonlocal operator} that encapsulates the influence of the hydrodynamics, and we describe the approach of preconditioning with continuous operators. In Section \ref{AccuracySection}, we  establish the method as third-order accurate, based on careful analysis of the flow singularities, and we conduct a convergence study to confirm this result. In Section \ref{BenchmarkSection} we derive a new class of asymptotic solutions to benchmark the numerical method. In Section \ref{ResultsSection}, we present a variety of examples to demonstrate the method's utility in understanding flexible-wing propulsion, and we close with a discussion in Section \ref{SectionDiscussion}.

\section{Problem formulation}
\label{FormulationSection}

\subsection{Physical setup}

\begin{figure}
\begin{center}
\includegraphics[width = 0.85 \textwidth]{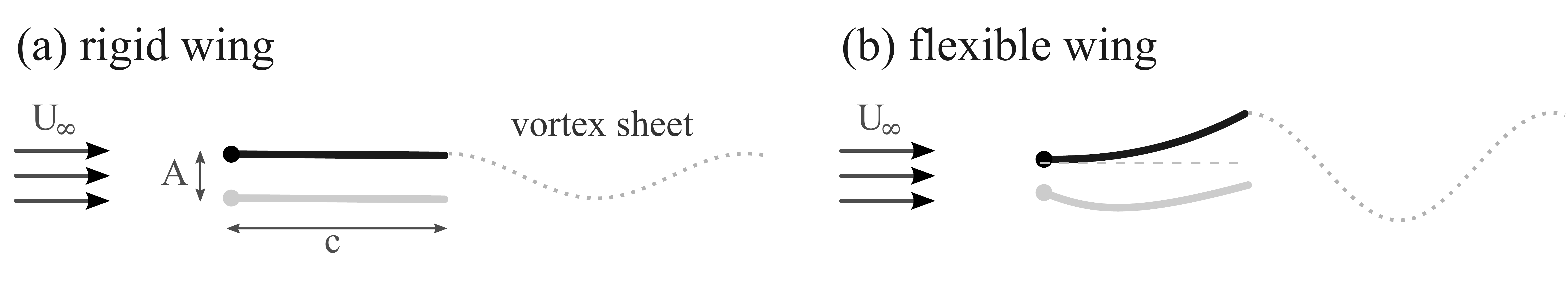}
\caption{ Schematic of a (a) rigid and (b) flexible wing, flapping in an oncoming flow.}
\label{schematic}
\end{center}
\end{figure}

As diagrammed in Fig.~\ref{schematic}, our physical setup consists of a thin wing flapping at small amplitude in a 2D fluid with free-stream velocity $\Udim$ perpendicular to the flapping motion. The fluid is considered inviscid, but we account for the viscous production of vorticity by allowing the wing to shed a vortex sheet from its trailing edge. The wing has chord length $\chord$ and thickness $\thick \ll \chord$, and is driven at the leading edge with characteristic amplitude $\amp$ and frequency $\freq$. This 2D setup can approximate a three-dimensional wing as long as the aspect ratio is sufficiently high, in which case $\width$ denotes the span and $\width \gg \chord$. The wing is allowed to be flexible, with an elastic modulus $E$ and mass per unit length $\mu$ that can vary with distance $x$ along the chord. We describe wing deformations in terms of the vertical displacement $h(x,t)$ as governed by the 1D beam equation,
\begin{equation}
\label{beamEq}
\mu(x) \ppd{h}{t} + \ppd{}{x} \left( \Ia E(x) \ppd{h}{x} \right) =  \loadDim(x,t) \, .
\end{equation}
Here, $\Ia = \width \thick^3/12$ the second moment of area of the wing's cross-section, and the external load $\loadDim$ is provided by the pressure difference across the wing
\begin{equation}
\loadDim(x,t) = \width \left( p^{-} - p^{+} \right) \, ,
\end{equation}
where $p^{-}$ and $p^{+}$ are the pressure distributions on the bottom and top surfaces of the wing respectively.

 	To determine these pressure distributions, we must solve for the surrounding fluid flow as governed by the 2D incompressible Euler equations
\begin{eqnarray}
\label{EulerEq}
\pd{\bu}{t} + \left( \bu \cdot \grad \right) \bu &=& - \frac{1}{\rho} \grad p \, , \\
\label{incomp0}
\grad \cdot \bu &=& 0 \, ,
\end{eqnarray}
Here, $\bu$ is the velocity field, $p$ is the pressure field, and $\rho$ is the fluid density (mass per unit volume). The flow is subject to no-penetration conditions along the wing surface, and the Kutta condition is enforced at the trailing edge to allow vorticity production.

\subsection{Dimensionless formulation}
\label{NondimSection}

To nondimensionalize, we choose the half-chord $\chord/2$ and the flapping period $1/\freq$ as the characteristic length and time scales respectively (giving a characteristic velocity of $\chord \freq/2$). After scaling on these values, Eq.~(\ref{beamEq}) becomes
\begin{equation}
\label{beamND}
2 \Mrat (x) \, \ppd{h}{t} + \frac{8 \pi^2}{3 \sigma^2} \, \ppd{}{x} \left( \Stiff(x) \ppd{h}{x} \right) = \load(x,t) \, ,
\end{equation}
where $x$, $t$, and $h$ are now understood to be dimensionless, with $x=-1$ representing the wing's leading edge and $x=1$ the trailing edge. The dimensionless hydrodynamic load $\load$ is given by
\begin{equation}
\load(x,t) = \frac{4 \left( p^{-} - p^{+} \right)} {\rho \chord^2 \freq^2} \, , \quad \forxc \, ,
\end{equation}
and we have introduced the following dimensionless parameters
\begin{equation}
\label{ParamEq}
\sigma = \frac{\pi \chord \freq}{\Udim} \, , \qquad
\Mrat(x) = \frac{\mu(x)}{\rho \width \chord} \, , \qquad
\Stiff(x) = \frac{E(x) \thick^3}{\rho \Udim^2 \chord^3} \, .
\end{equation}
Here, $\sigma$ is the reduced driving frequency, $\Mrat$ measures the ratio of solid-to-fluid inertia, and $\Stiff$ is the dimensionless wing stiffness (the ratio of elastic-to-fluid forces). We allow both $\Mrat$ and $\Stiff$ to vary with $x$, reflecting variations in the wing density and rigidity respectively. For convenience in physical applications, the wing's mass-per-unit-length can be expressed as $\mu = \rho_s \width \thick$, where $\rho_s$ is the density of the solid material comprising the wing (mass per unit volume). Although this paper focuses on 2D simulations, we have defined pressure and density as three-dimensional physical quantities (force-per-unit-area and mass-per-unit-volume respectively) for ease in applying the theory to real wings/fins of high aspect ratio.

\subsection{Small-amplitude linearization}
To treat the fluid flow, we write the incompressible Euler equations (\ref{EulerEq})--(\ref{incomp0}) in dimensionless variables and linearize in small-amplitude to obtain
\begin{eqnarray}
\label{linEuler}
\dtOpp \bu &=& \grad \paccel \, , \\
\label{incomp}
\grad \cdot \bu &=& 0 \, .
\end{eqnarray}
Here, $\bu = (\UFS + u, v)$ is the  velocity field relative to the dimensionless free-stream value $\UFS = 2 \Udim/ (\chord \freq)$. Note that $\UFS = 2 \pi / \sigma$, and so the number of independent dimensionless variables remains unchanged. The above linearization is valid as long as $\amp/\chord \ll 1$ and $\freq \amp / \Udim \ll 1$ \cite{Wu1961, Moore2014}.
The function $\paccel$ is known as the {\em Prandtl acceleration potential} and is given by
\begin{equation}
\paccel(x,y,t) = \frac{4 \left( \pinf - p \right) }{\rho \chord^2 \freq^2 } \, .
\end{equation}
Thus, $\paccel$ has the physical interpretation of a {\em negative, dimensionless pressure field}. In terms of $\paccel$, the dimensionless hydrodynamic load is given by
\begin{equation}
\label{hload}
\load(x,t) = \paccel(x,0^+,t) - \paccel(x,0^-,t) \, , \quad \forxc \, .
\end{equation}
When linearized in small-amplitude, the no-penetration boundary condition on the wing $\wing = \{ x \in [-1,1], \, y=0 \}$ becomes
\begin{equation}
\label{wingBC}
\left. v \right|_{\wing} = \dtOpp h \, .
\end{equation}
Meanwhile, the Kutta condition requires the velocity at the trailing edge to be finite
\begin{equation}
\label{Kutta}
\left. \abs{v} \, \right|_{(x,y)=(1,0)} < \infty \, ,
\end{equation}
thus setting the rate of vorticity production.

\section{Semi-analytical calculation of the fluid flow}
\label{FluidSection}

In this section, we derive a class of semi-analytical solutions for the flow governed by Eqs.~(\ref{linEuler})--(\ref{incomp}) and (\ref{wingBC})--(\ref{Kutta}) under the assumption that the wing kinematics are {\em given}. Later, we will couple these solutions to unknown kinematics through Eq.~(\ref{beamND}) to {\em solve} for the emergent wing deformations. The solutions derived here can also be found in the literature \cite{Wu1961, Moore2014}, and so we provide only a brief recap, with emphasis given to the parts of the calculation that are most relevant for our numerical method.

To begin, suppose the wing kinematics are harmonic in time with a spatial component represented by a Chebyshev series,
\begin{align}
\label{hDecomp}
& h(x,t) = \ejt \eta(x) \, , \\
\label{etaExp}
& \eta(x) = \sump{k=0}{\infty} \hat{\eta}_k T_k(x) 
\coloneqq \frac{1}{2} \hat{\eta}_0 + \sum_{k=1}^{\infty} \hat{\eta}_k T_k(x) \, .
\end{align}
Here, $j = \sqrt{-1}$ and it is implied that the real part in $j$ should be taken in Eq.~(\ref{hDecomp}); $\eta(x) = \eta_R(x) + j \eta_I(x)$ is a complex-valued function, whose real and imaginary parts together determine the motion of the wing;
 $T_k(x) = \cos (k \arccos x)$ is the Chebyshev polynomial of degree $k$, and the sum-prime notation indicates that the $k=0$ term should be multiplied by $1/2$ as shown above.

	To solve for the emergent flow, we will identify 2D physical space $(x,y)$ with the complex plane $z = x+iy$. Taking the divergence of Eq.~(\ref{linEuler}) and using incompressibility shows that $\paccel$ is a harmonic function, which implies the existence of a harmonic conjugate $\psi$. Their combination produces the complex acceleration potential, $\compot(z,t) = \paccel + i \psi$, which is an analytic function of $z$. Using Eq.~(\ref{linEuler}), $\compot$ relates to the complex velocity field $w = u-iv$ through
\begin{equation}
\label{compvel}
\pd{\compot}{z} = \pd{w}{t} + \UFS \pd{w}{z}
\end{equation}
We conformally map the physical domain (the $z$-plane) to the exterior of the unit disk (the $\zeta$-plane) through
\begin{equation}
\label{cmap}
z = \frac{1}{2} \left( \zeta + \frac{1}{\zeta} \right) \, ,
\end{equation}
where the wing surface maps to the unit circle. In the $\zeta$-plane, the complex potential can be represented by a multipole expansion
\begin{equation}
\label{mpole}
\compot = \paccel + i \psi = i \ejt \left( \frac{\co_0}{\zeta+1} + \sum_{k=1}^{\infty} \frac{\co_k}{\zeta^{k}} \right) \, ,
\end{equation}
where the coefficients $\co_k$ are unknowns that will depend on the wing kinematics. In this expansion, a singularity is allowed at $\zeta=-1$, corresponding to the wing's leading edge, but not at $\zeta=1$, the trailing edge, in order to satisfy the Kutta condition \cite{Wu1961, Moore2014}. The coefficients $\co_k$ are real with respect to $i$ (due to the up-down symmetry of the physical setup), but can be complex with respect to $j$ to reflect temporal phases \cite{Wu1961, Moore2014}. Evaluating the real and imaginary parts of Eq.~(\ref{mpole}) on the unit circle $ \ucirc = \{ \zeta = e^{i \theta} \}$ yields the following expansions
\begin{align}
\label{phitheta}
& \left. \paccel \right|_{\ucirc} = \ejt \left( \frac{1}{2} \co_0 \tan \frac{\theta}{2} + 
							\sum_{k=1}^{\infty} \co_k \sin{k \theta} \right) \, , \\
\label{psitheta}
& \left. \psi \right|_{\ucirc} =  \ejt \sump{k=0}{\infty} \co_k \cos{k \theta} \, .
\end{align}

	Now consider evaluating these functions on the wing, with the top surface denoted $\wing^+ = \{ x \in [-1,1], y=0^{+} \}$ and the bottom $\wing^- = \{ x \in [-1,1], y=0^{-} \}$. Under transformation (\ref{cmap}), a point $x$ on the wing maps to a point $\zeta = e^{i \theta}$ on the unit circle via $x = \cos \theta$. Since $\psi$ is even in $\theta$, it will have the same values on both sides. However, $\paccel$ is odd in $\theta$ and thus will suffer a jump discontinuity. We separate each of these functions into temporal and spatial components to get
\begin{align}
\label{psiDecomp}
& \left. \psi \right|_{\wing} \,\, = \ejt \Psi(x) \, , \\
\label{phiDecomp}
& \left. \paccel \right|_{\wing^{\pm}} = \ejt \Phi^{\pm}(x) \, ,
\end{align}
The cosine series in Eq.~(\ref{psitheta}) gives a Chebyshev series for $\Psi(x)$
\begin{equation}
\label{PsiExp}
\Psi(x) = \sump{k=0}{\infty} \co_k T_k(x) \, ,
\end{equation}
Meanwhile, Eq.~(\ref{phitheta}) produces a series for $\Phi^{\pm}(x)$ that can be expressed as
\begin{equation}
\label{PhiExp}
\Phi^{\pm}(x) = \pm \frac{\co_0}{2} \singterm \pm \sum_{k=1}^{\infty} \co_k \sin {k \theta} \, ,
\end{equation}
where $\theta = \arccos x$, and we have used the identity $\tan(\theta/2) = \sqrt{(1-x)/(1+x)}$.

	Taking the imaginary part of Eq.~(\ref{compvel}), evaluating on the wing surface, and using boundary condition (\ref{wingBC}), yields
\begin{equation}
\label{wingBC2}
\left. \pd{\psi}{x} \right|_{\wing}  = -\dtOpp^2 h \, .
\end{equation}
Using Eqs.~(\ref{hDecomp}) and (\ref{psiDecomp}), the above equation simplifies to
\begin{equation}
\label{DPsi}
\Dx \Psi = -(2 \pi j + U \Dx)^2 \eta \, ,
\end{equation}
where $\Dx = d / dx$. Given the kinematics $\eta$, this equation allows us to solve for $\Psi$ up to a constant of integration. Explicit formulas relating the coefficients $\etah_k$ and $\co_k$ can be found in Wu (1961) \cite{Wu1961}. However, we leave Eq.~(\ref{DPsi}) in its current form due to the conceptual simplicity in applying fast Chebyshev differentiation/integration routines that will be introduced later.

	To complete the calculation of $\Psi$, it is necessary to determine the constant of integration $\co_0$. This value has been calculated in previous studies \cite{Kussner41, Wu1961, Li2015}, and  we will simply quote the result here. We must first decompose the vertical velocity into temporal and spatial components
\begin{align}
& \left. v \right|_{\wing} = \ejt V(x) \, , \\
\label{VExp}
& V(x) = \sump{k=0}{\infty} \, \vn_k T_k(x) \, ,
\end{align}
Then, boundary condition~(\ref{wingBC}) can be written as
\begin{equation}
\label{VEq}
V = ( 2 \pi j + U \Dx ) \eta
\end{equation}
Given $\eta(x)$, this equation allows us to determine $V(x)$. The coefficient $\co_0$ can then be expressed as \cite{Kussner41, Wu1961, Li2015}
\begin{equation}
\label{a0Eq}
\co_0 = -\UFS \theofun(\sigma) (\vn_0 + \vn_1) + \UFS \vn_1 \, .
\end{equation}
Here, $\theofun(\sigma)$ is the Theodorsen function given by $\theofun(\sigma) = K_1(j \sigma) / ( K_0(j\sigma) + K_1(j\sigma) )$, where $K_0$ and $K_1$ are modified Bessel functions of the second kind.

	With all of the coefficients $\co_k$ known, Eqs.~(\ref{phiDecomp}) and (\ref{PhiExp}) give the pressure distributions along the top and bottom surfaces of the wing. These expressions can then be inserted into Eq.~(\ref{hload}) to give the hydrodynamic load as
\begin{align}
\label{loadDecomp}
& \load(x,t) = \ejt \loadx(x) \, , \\
\label{loadxEq}
& \loadx(x) = \co_0 \singterm  + 2 \sinsum \, ,
\end{align}
where $\theta = \arccos x$.

In summary, for wing kinematics given by $h(x,t) = \ejt \eta(x)$, we use Eqs.~(\ref{etaExp}), (\ref{PsiExp}), and (\ref{DPsi}) to calculate $\co_k$ for $k=1,2,3,\dots$ and Eqs.~(\ref{VExp})--(\ref{a0Eq}) to calculate $\co_0$. The hydrodynamic load is then given by Eqs.~(\ref{loadDecomp}) and (\ref{loadxEq}). Since the coefficients in this series depend on the kinematics, the hydrodynamic load can be regarded as an operator that acts on the function $\eta(x)$, i.e.~$\loadx = \loadx[\eta]$. Furthermore, Eqs.~(\ref{DPsi}), (\ref{VEq}), and (\ref{a0Eq}) all depend {\em linearly} on $\eta$, and thus $\loadx[\eta]$ is a {\em linear} operator.

{\newa We note that it is possible to use these flow solutions to compute trailing vorticity as a post-processing step \cite{Wu1961}. This calculation would reveal a flat vortex sheet lying along $\{y=0, \, x>1\}$ with periodically varying vorticity. We remind the reader, though, that vorticity is not needed to determine the wing kinematics nor the associated hydrodynamic forces. Rather, our method features {\em pressure} as the primary variable of interest.}

\section{Numerical method for the wing's bending motion}
\label{BendingSection}

\subsection{The boundary-value problem with nonlocal operator}
We now turn our attention to solving for the emergent deformations of a flexible wing. Note that, by the time-harmonic assumption in Eq.~(\ref{hDecomp}), the complex-valued function $\eta(x) = \eta_R(x) + j \eta_I(x)$ {\em completely determines the wing kinematics} through
\begin{equation}
\label{heta}
h(x,t) = \eta_R(x) \cos(2 \pi t) - \eta_I(x) \sin(2 \pi t) \, .
\end{equation}
 It is therefore our goal to solve for $\eta(x)$. Inserting Eqs.~(\ref{hDecomp}) and (\ref{loadDecomp}) into Eq.~(\ref{beamND}) gives the fourth-order  differential equation for $\eta(x)$
\begin{equation}
\label{etaODE}
\Dx^2 \left( \alpha(x) \Dx^2 \eta \right) - \beta(x) \eta = \loadx[\eta](x) \, , 
\quad \forxo \, .
\end{equation}
We have introduced the variable coefficients
\begin{equation}
\label{alphabeta}
\alpha(x) = \frac{8 \pi^2}{3 \sigma^2} \Stiff(x) \, , \quad \beta(x) = 8 \pi^2 \Mrat(x) \, ,
\end{equation}
which depend on the wing stiffness and mass distributions respectively. The wing is subject to imposed heaving and pitching at the leading edge ($x=-1$), while the trailing edge ($x=1$) is a free end. This arrangement gives boundary conditions
\begin{align}
\label{etaBC1}
& \eta(-1) = \etaLE \, , \quad \eta'(-1) = \etaLE' \, , \\
\label{etaBC2}
& \eta''(1) = 0 \, , \qquad \eta'''(1) = 0 \, .
\end{align}
Here, $\etaLE$ and $\etaLE'$ represent the leading-edge heaving and pitching respectively. Equations (\ref{etaODE})--(\ref{etaBC2}) constitute a boundary value problem (BVP) for $\eta(x)$.

At first glance, the hydrodynamic load, $\load$, appears to enter the beam equation (\ref{beamND}) as an inhomogeneous term. However, Section \ref{FluidSection} shows that the load actually depends on the wing kinematics. We therefore use the notation $\loadx[\eta](x)$ in Eq.~(\ref{etaODE}) to remind the reader that the (spatially varying) hydrodynamic load comes from a linear operator that acts on $\eta(x)$. As such, Eq.~(\ref{etaODE}) is a {\em linear, homogeneous} differential equation with {\em nonlocal term} $\loadx[\eta]$.

	The results from Section \ref{FluidSection} offer an efficient way to calculate the action of the operator $\loadx[\eta]$ for given kinematics. However, these gains would be lost if we were to explicitly construct the matrix corresponding to $\loadx[\eta]$ in order to solve system (\ref{etaODE})--(\ref{etaBC2}) for unknown kinematics. We will therefore solve the system with a Krylov-subspace iterative method, since then we need only apply $\loadx[\eta]$ {\em forward} in the matrix-free fashion suggested by Section \ref{FluidSection}. Two problems result from applying an iterative method to system (\ref{etaODE})--(\ref{etaBC2}) in its current form. First, since Eq.~(\ref{etaODE}) involves a fourth-order derivative, the corresponding discretized system is poorly conditioned, and so an iterative method would require many iterations to converge. Second, the first term on the right-hand-side of Eq.~(\ref{loadxEq}) is singular at $x=-1$, which would even more severely impede convergence and accuracy. Both of these problems can be eliminated by preconditioning. In the next section, we discuss a method to precondition system (\ref{etaODE})--(\ref{etaBC2}) in its {\em continuous} form, rather than working with the corresponding matrix.

\begin{figure}
\begin{center}
\includegraphics[width = 0.65 \textwidth]{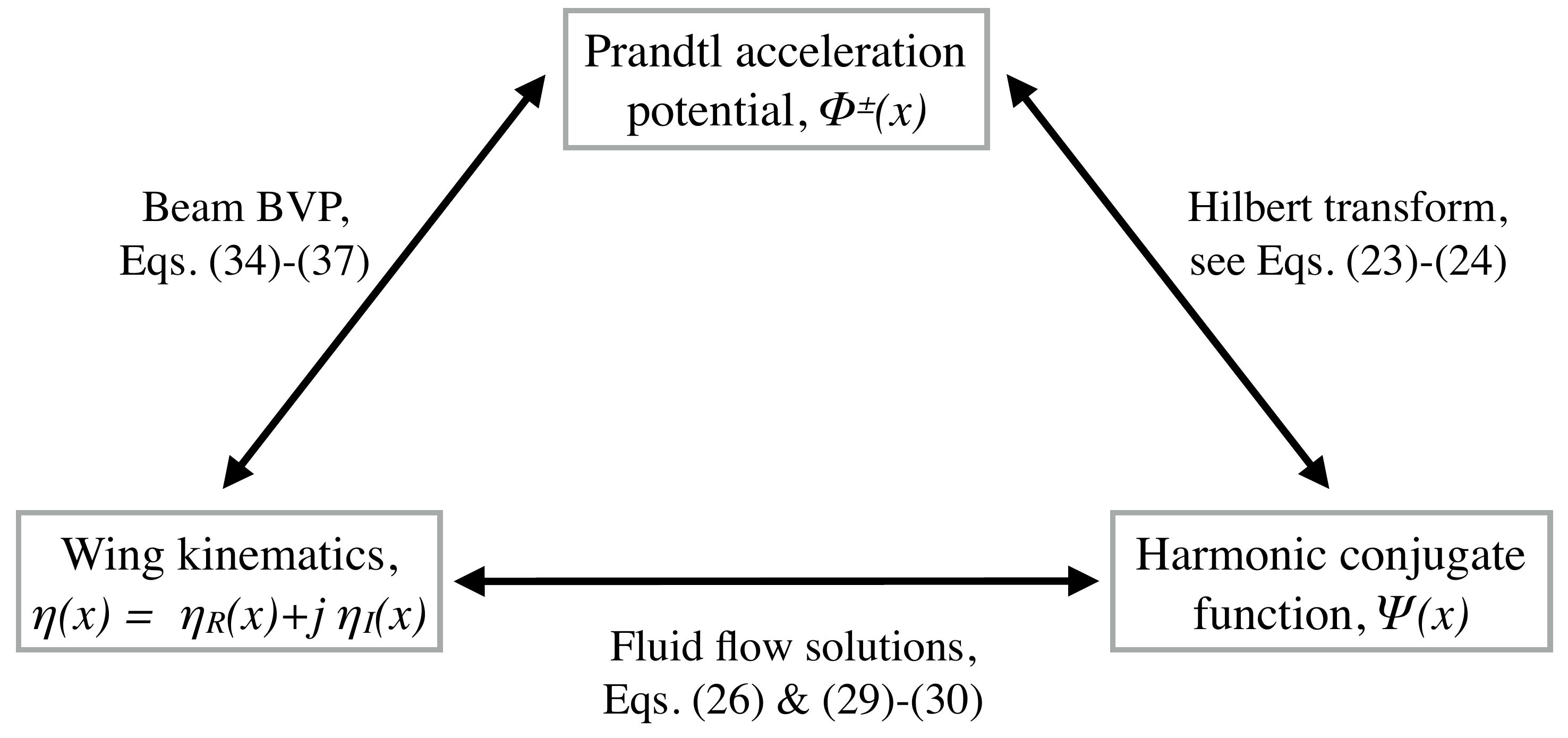}
\caption{ Mathematical structure of the problem: To solve for the emergent wing kinematics, we must simultaneously determine $\eta(x)$, $\Phi^{\pm}(x)$, and $\Psi(x)$. Here, we diagram the relationships between each of these functions.
}
\label{RHprob}
\end{center}
\end{figure}
 
	Before that discussion, we briefly comment on the overall mathematical structure of our problem, which is summarized schematically in  Fig.~\ref{RHprob}. The main unknowns at play are the wing kinematics, $\eta(x)$, the Prandtl acceleration potential, $\paccel$, and its harmonic conjugate, $\psi$. More specifically, we are interested in the boundary values of the latter two, $\Phi^{\pm}(x)$ and $\Psi(x)$ respectively. BVP (\ref{etaODE})--(\ref{etaBC2}) links $\eta(x)$ to the hydrodynamic load $\loadx(x)$, which is obtained directly from $\Phi^{\pm}(x)$ (see Eq.~(\ref{hload})). Meanwhile, the results from Section \ref{FluidSection} relate $\eta(x)$ and $\Psi(x)$ (see Eqs.~(\ref{DPsi}), (\ref{VEq}), and (\ref{a0Eq})). To close the loop, $\paccel$ and $\psi$ are harmonic conjugates and so their boundary values are related by a type of Hilbert transform\footnotemark[1] \cite{Alben2008, Shelley2011}.  As such, our problem of simultaneously determining $\eta(x)$, $\Phi^{\pm}(x)$, and $\Psi(x)$ is closely related to the celebrated Riemann-Hilbert problem of determining a harmonic function from boundary-data of its conjugate function \cite{Wu1971, Ablowitz2003, Its2003, Olver2014, Huang2015, MooreCPAM}. In our case, though, the two harmonic functions are also constrained by differential equations involving a common function, $\eta(x)$, which itself is unknown. As such, our reformulation of the original PDE system might be described as a Riemann-Hilbert problem with ODE constraints\footnotemark[2].

\footnotetext[1]{in this case, a Hilbert transform over the interval $x \in (-1,1)$ with singularity at $x=-1$.}
\footnotetext[2]{or equivalently an ODE system with Riemann-Hilbert constraint; it matters not which is deemed the problem and which the constraint.}

	Finally, the relationship between $\Phi^{\pm}(x)$ and $\Psi(x)$ can be seen directly through Eqs.~(\ref{PsiExp}) and (\ref{PhiExp}), as these two expansions are linked by the coefficients $\co_n$. In what follows, we will perform the Hilbert transform between $\Phi^{\pm}(x)$ and $\Psi(x)$ by going through the coefficients $\co_n$ as an intermediary. These calculations can be accomplished in a straightforward and efficient manner by using discrete cosine and sine transforms (DCT and DST respectively).

\subsection{Preconditioning with continuous operators}

We now discuss how to precondition system (\ref{etaODE})--(\ref{etaBC2}) using {\em continuous} operators. {\newb First, focusing on the highest-order derivative present in Eq.~(\ref{etaODE})}, we define the operator $\Popp$ to act on a function $u(x)$ via
\begin{equation}
\label{Popp}
\Popp [u] = \Dx^2 \left( \alpha(x) \Dx^2 u \right) \, .
\end{equation}
For a (sufficiently smooth) function $v(x)$, we define $u = \Pinv[v]$ as the solution to\footnotemark[3]
\begin{align}
\label{Puv}
& \Popp u = v, \qquad \forxo \, \\
\label{HomogBCs}
& u(-1) = u'(-1) = u''(1) = u'''(1) = 0 \, ,
\end{align}
That is, $\Pinv$ inverts $\Popp$ subject to certain {\em homogeneous} boundary conditions. If a function $w(x)$ satisfies conditions (\ref{HomogBCs}), then $\Pinv \Popp w = \Popp \Pinv w = w$. However, if $w$ does not satisfy these boundary conditions, then $\Pinv \Popp w \ne w$ in general.

\footnotetext[3]{
Here, $u$ and $v$ do not correspond to fluid velocity components as they do in Section \ref{FluidSection}.}

We precondition Eq.~(\ref{etaODE}) by applying $\Pinv$ on the left,
\begin{equation}
\label{pcEq0}
\Pinv \Popp [\eta] - \Pinv \left( \beta(x) \eta + \loadx[\eta] \right) = 0 \, .
\end{equation}
The solution that we seek satisfies {\em inhomogeneous} boundary conditions (\ref{etaBC1})--(\ref{etaBC2}) and therefore $\Pinv \Popp [\eta] \ne \eta$. To proceed, we introduce the linear function
\begin{equation}
\label{etalin}
\etalin(x) = \etaLE + \etaLE' (x+1) \, ,
\end{equation}
which satisfies $\Popp[\etalin] = 0$ and boundary conditions (\ref{etaBC1})--(\ref{etaBC2}). Consequently, $\eta(x)-\etalin(x)$ satisfies homogeneous boundary conditions (\ref{HomogBCs}), implying that $\Pinv \Popp [\eta-\etalin] = \eta-\etalin$. We make this substitution in Eq.~(\ref{pcEq0}) and rearrange to get
\begin{equation}
\label{pcEq}
\eta - \Pinv \left( \beta(x) \eta + \loadx[\eta] \right) = \etalin(x) \, ,
\end{equation}
This is the preconditioned form of system (\ref{etaODE})--(\ref{etaBC2}). In fact, since $\Pinv$ is an integral operator, {\newb Eq.~(\ref{pcEq}) can be regarded as an integral-reformation of the original system \cite{Greengard1991, GreenRokh1991}. }

The problem of the singular term in Eq.~(\ref{loadxEq}) remains. Here, we will apply $\Pinv$ {\em semi-analytically} in order to remove the singularity. We first decompose the hydrodynamic load into singular and regular components, $\loadx(x) = \co_0 \loadsing(x) + \loadreg(x)$, where
\begin{equation}
\label{loadDecomp2}
\loadsing(x) = \singterm \, , \qquad
\loadreg(x) = 2 \sinsum \, .
\end{equation}
We next define $\eta_s(x) = \Pinv \left[ \loadsing \right]$, or, more explicitly, $\eta_s(x)$  solves
\begin{align}
\label{singODE0}
& \Dx^2 \left( \alpha(x) \Dx^2 \eta_s \right) = \singterm \quad \forxo \, , \\
& \eta_s(-1) = \eta_s'(-1) = \eta_s''(1) = \eta_s'''(1) = 0 \, .
\end{align}
Integrating twice, while satisfying $\eta_s''(1) = \eta_s'''(1) = 0$, gives a second-order ODE for $
\eta_s(x)$,
\begin{align}
\label{singODE}
& \Dx^2 \eta_s = \frac{1}{ 2 \alpha(x)} \left( (2+x) \sqrt{1-x^2} - (1+2x) \arccos x \right) \, , \\
\label{singBCs}
& \eta_s(-1) = \eta_s'(-1) = 0 \, .
\end{align}
Here, we have removed the singularity in Eq.~(\ref{singODE0}) through antidifferentiation. The right-hand side of Eq.~(\ref{singODE}) is continuously differentiable, though its second derivative is singular.

Returning to Eq.~(\ref{pcEq}), we separate out the singular part to obtain
\begin{equation}
\label{pcdsEq}
\eta - \co_0[\eta] \, \eta_s(x) - \Pinv \left( \beta(x) \, \eta + \loadreg[\eta](x) \right) = \etalin(x) \, .
\end{equation}
Equation (\ref{pcdsEq}) is therefore the {\em preconditioned} and {\em desingularized} form of the original BVP (\ref{etaODE})--(\ref{etaBC2}). Notice that boundary conditions (\ref{etaBC1})--(\ref{etaBC2}) are implied by the right-hand-side of Eq.~(\ref{pcdsEq}) and the definition of $\Pinv$.

{\newb
We now aim to solve Eq.~(\ref{pcdsEq}) for $\eta(x)$ with an iterative method, while always computing the nonlocal terms $\co_0[\eta]$ and $\loadreg[\eta](x)$ in the matrix-free fashion offered by Section \ref{FluidSection}. A few observations are in order. First, the small-amplitude framework from Section \ref{FluidSection} already places the main quantities of interest in Chebyshev space, which makes Chebyshev-spectral methods \cite{Mason2002, Gil2007} the natural choice for handling the differential/integral operations involved in Eq.~(\ref{pcdsEq}). These methods will be described in the next section. Second, thanks to the preconditioning, the majority of the needed operations are of the integral type, which are numerically stable and easy to perform in Chebyshev space.  It is only the computation of the nonlocal operator that requires numerical differentiation, specifically Eqs.~(\ref{DPsi}) and (\ref{VEq}). 

Before discussing the Chebyshev methods, we comment that we could have just as easily chosen to {\em right}-precondition our system with $\Popp$, rather than left-precondition. Right-preconditioning would be equivalent to a traditional integral-reformulation \cite{Greengard1991}, and offers some advantage in that it could eliminate the need to perform any numerical differentiation whatsoever. However, the main reason we chose to left-precondition is the ability to treat the singular term analytically. Removing this singularity in the right-conditioned system appears to be less straightforward.
}

\subsection{Chebyshev spectral-collocation routines}
\label{ChebySection}

{\newb
As mentioned above, the majority of the operations needed to iterate Eq.~(\ref{pcdsEq}) are of the integral type, e.g.~$\Dinv$ and $\Pinv$, which are numerically stable operations \cite{Greengard1991}. Only in computing the nonlocal terms do we need to perform numerical differentiation, specifically in Eqs.~(\ref{DPsi}) and (\ref{VEq}). Furthermore, we need only compute {\em first}-order derivatives in each case ($\Dinv$ can be applied to both sides of Eq.~(\ref{DPsi})). For this, we will use the transform-recursive method \cite{Don1995}, which enables fast numerical differentiation via recursive formulas. This method can be numerically unstable for very large $N$, as errors can be amplified by a factor of $\BigO{N^2}$ \cite{Don1995, Greengard1991}. Thanks to its high accuracy, though, our required $N$-values will be moderate (typically less than 1000), for which the growth of roundoff error poses no threat. Even for $N$-values on the order of 16,000, we observe no such numerical instabilities in the convergence tests conducted in Section \ref{ConvTest}. In the following, we perform all integration/differentiation in spectral space and all multiplications in physical space, i.e.~a pseudo-spectral method. \\
} 

\noindent
{\bf Chebyshev collocation and the DCT: }
We first introduce the (interior) Gauss-Chebyshev points \cite{Mason2002, Gil2007}
\begin{align}
& x_n = \cos \theta_n \, , \\
& \theta_n = \frac{\pi (2n+1)}{2(N+1)} \qquad \text{for } n=0, 1, \ldots , N \, .
\end{align}
As shown here, the collocation points $x_n$ are related to an equispaced  $\theta$-grid. 
Consider a function $f(x)$ interpolated at these points by polynomial $p_{N}(x)$ of degree $N$, which itself is expressed as a Chebyshev sum,
 \begin{align}
& f(x_n) = p_N(x_n)  \quad \text{for } n=0,1, \ldots , N \, , \\
\label{ChebPoly}
& p_N(x) = \sump{k=0}{N} \pco_k T_k(x) \, .
\end{align}
Transforming to the $\theta$-grid produces a cosine series for the interpolation values
\begin{equation}
f(x_n) = \sump{k=0}{N} \pco_k \cos (k \theta_n)  \quad \text{for } n=0,1, \ldots , N \, .
\end{equation}
This reformulation allows one to use the discrete cosine transform (DCT) to transform between a function's collocation values, $f_n = f(x_n)$, and the Chebyshev coefficients, $\pco_k$. With the DCT, the transform can be performed efficiently with $\BigO{N \log N}$ operations  \cite{Don1995, Mason2002, Gil2007}.

\vsp{1}
{\bf Fast spectral integration and differentiation: }
To describe the fast integration and differentiation routines, we begin with the {\em Chebyshev recurrence relation} \cite{Don1995, Mason2002, Gil2007}
\begin{equation}
\label{ChebyRecur}
2 T_k(x) = \frac{1}{k+1} T_{k+1}'(x) - \frac{1}{k-1} T_{k-1}'(x) \, .
\end{equation}

Antidifferentiating Eq.~(\ref{ChebPoly}) and using the recurrence relation produces an {\em explicit} formula for the coefficients of the antiderivative,
\begin{align}
\label{intpoly}
& \Dinv  p_N(x) = \sump{k=0}{N+1} \Pco_k T_k(x) \, , \\
\label{fastint}
&  \Pco_k = \frac{1}{2k} \left( \pco_{k-1} - \pco_{k+1} \right) \qquad \text{for } k=1, 2, \ldots , N+1 \, .
\end{align}
Here, $\Dinv $ denotes an antiderivative and, naturally, $\Pco_0$ is a free constant of  integration.

Meanwhile, differentiating Eq.~(\ref{ChebPoly}) and using Eq.~(\ref{ChebyRecur}) produces a {\em recurrence relation} for the  coefficients of the derivative,
\begin{align}
& \Dx p_N(x) = \sump{k=0}{N} \ppco_k T_k(x) \, , \\
&  \ppco_{N+1} = \ppco_{N} = 0 \, , \\ 
& \ppco_k = \ppco_{k+2} + 2(k+1) \pco_{k+1}	\quad \text{for } k=N-1, N-2, \ldots , 0 \, .
\end{align}
The coefficients $\ppco_k$ can therefore be computed with only $\BigO{N}$ operations. 

In summary, given the coefficients $\pco_k$ of the interpolating polynomial $p(x)$, we can compute the coefficients of the antiderivative and derivative  with $\BigO{N}$ operations. Transforming back to physical space ($\BigO{N \log N}$ operations) produces an approximation to  $\Dinv  f$  and $\Dx f$ at the collocation points.

\vsp{1}
{\bf Evaluation at the endpoints:}
To enforce boundary conditions, we must be able to evaluate functions at the endpoints $x=\pm 1$, which are not part of the collocation grid. For this, we use the special formulas
\begin{equation}
\label{fpm1}
p_N(\pm 1) = \sump{k=0}{N} (\pm1)^k \pco_k \, .
\end{equation}

\vsp{1}
{\bf Fast BVP solvers and fast preconditioning:}
The main challenge posed by Eq.~(\ref{pcdsEq}) is applying the preconditioner $\Pinv$. Consider computing $u = \Pinv[v]$ for an arbitrary function $v(x)$, which is equivalent to solving BVP (\ref{Puv})--(\ref{HomogBCs}) for $u(x)$. This fourth-order BVP can be broken down into two second-order problems. The first is a BVP for an auxiliary function $w(x)$,
\begin{align}
\label{wODE}
& \Dx^2 w = v	\qquad \text{for } x \in (-1,1) \, , \\
\label{wBCs}
& w(1) = w'(1) = 0 \, .
\end{align}
The second is a BVP is for the desired function $u(x)$,
\begin{align}
\label{uODE}
& \Dx^2 u = w/\alpha	\qquad \text{for } x \in (-1,1) \, , \\
\label{uBCs}
& u(-1) = u'(-1) = 0 \, .
\end{align}
Notice that the boundary conditions $u''(1) = u'''(1) = 0$ are automatically satisfied as a result of  Eqs.~(\ref{wBCs}) and (\ref{uODE}).

These two BVPs share a common form and can therefore be solved with the same algorithm. The common form of BVPs (\ref{wODE})--(\ref{wBCs}) and (\ref{uODE})--(\ref{uBCs}) is
\begin{align}
\label{ODE2nd}
& \Dx^2 f = g	\qquad \text{for } x \in (-1,1) \, , \\
\label{BC2nd}
& f(x_{end}) = f'(x_{end}) = 0 \, .
\end{align}
where $x_{end}$ is either 1 or -1. {\newb We now give an algorithm to solve BVP (\ref{ODE2nd})--(\ref{BC2nd}) in spectral space:}

\begin{algorithm}[H]
\caption{Solve the second-order BVP (\ref{ODE2nd})--(\ref{BC2nd}) {\newb in spectral space.}}
\label{BVP2Algo}
\begin{algorithmic}
\State {\bf Input}: {\newb The $N+1$ Chebyshev coefficients of $g(x)$, denoted $\hat{g}_k$ for $k=0,1,\ldots,N$. }
\State 1. Use Eqs.~(\ref{intpoly})--(\ref{fastint}) to calculate a provisional antiderivative $\Dinv g$ (in spectral space).
\State 2. Use Eq.~(\ref{fpm1}) to evaluate $\Dinv g (x_{end})$ and then subtract this value in order to enforce $\Dinv g (x_{end}) = 0$.
\State 3. Repeat steps 1 and 2 (while remaining in spectral space)  to compute a second antiderivative $f = \Dx^{-2}g$, such that $f(x_{end}) = 0$, and return as output.
\end{algorithmic}
\end{algorithm}

\noindent
{\newb This algorithm allows one to solve BVPs of the special form (\ref{ODE2nd})--(\ref{BC2nd}) in spectral space with only $\BigO{N}$ operations. The physical solution can be recovered, if desired, by simply transforming back to physical space at an $\BigO{N \log N}$ cost.
Now, to compute $\Pinv[v]$, we first apply Algorithm \ref{BVP2Algo} to solve BVP  (\ref{wODE})--(\ref{wBCs}), then transform to physical space to compute $w(x)/\alpha(x)$, then transform back to spectral space and apply Algorithm \ref{BVP2Algo} once more to solve BVP (\ref{uODE})--(\ref{uBCs}). }
We therefore have the ability to precondition with $\BigO{N \log N}$ operations. Lastly, BVP (\ref{singODE})--(\ref{singBCs}) for the singular component, $\eta_s(x)$, also takes the form of Eqs.~(\ref{ODE2nd})--(\ref{BC2nd}) and can therefore be solved with the same algorithm.

\subsection{The main algorithm to solve for the wing kinematics}

With the Chebyshev routines in hand, we now discuss how to solve Eq.~(\ref{pcdsEq}) for the emergent wing kinematics, $\eta(x)$. To collect ideas, we define the linear operator $\Lopp$, which acts on a (complex-valued) trial function $\trialf(x)$ via
\begin{equation}
\label{Lopp}
\Lopp [\trialf] \coloneqq \trialf - \co_0[\trialf] \, \eta_s(x) - \Pinv \left( \beta(x) \, \trialf + \loadreg[\trialf](x) \right) \, .
\end{equation}
Recall that $\co_0[\trialf]$ and $\loadreg[\trialf](x)$ are themselves operators that act on $\trialf(x)$. With the above definition, the {\em preconditioned, desingularized} problem (\ref{pcdsEq}) can be compactly expressed as
\begin{equation}
\label{etaLinEq}
\Lopp [\eta] = \etalin(x) \, ,
\end{equation}
where $\etalin(x)$ is the linear function defined in Eq.~(\ref{etalin}). 

We aim to solve Eq.~(\ref{etaLinEq}) using the (matrix-free) generalized minimal residual method (GMRES) \cite{Saad1986, Evans1994}. We therefore must have an algorithm to apply $\Lopp$ to a given trial function $\trialf(x)$. Notice that $\Lopp$ depends on $\eta_s$, which itself is found by solving BVP (\ref{singODE})--(\ref{singBCs}). However, that BVP does not depend on $\trialf$, and therefore $\eta_s$ should be {\em precomputed} (using Algorithm \ref{BVP2Algo}). We now give the algorithm to apply $\Lopp$, assuming that $\eta_s$ has already been precomputed.  {\newb In this algorithm, $\trialf(x)$ and $\eta_s(x)$ are input in spectral space, the variable coefficients $\alpha(x)$ and $\beta(x)$ are input in physical space, and the result $\Lopp [\trialf]$ is output in spectral space.}

\begin{algorithm}[H]
\caption{Matrix-free application of the operator $\Lopp$ to a trial function $\trialf(x)$.}
\label{LAlgo}
\begin{algorithmic}

\State {\newb{\bf Main Inputs:} The $N+1$ Chebyshev coefficients of $\trialf(x)$ and of $\eta_s(x)$
\State {\bf Other Inputs:} The variable coefficients $\alpha(x)$ and $\beta(x)$ at the $N+1$ collocation points.
\State
\State \MyComment The objective of steps 1-4 is to compute the term 
$\Pinv \left( \beta(x) \, \trialf(x) + \loadreg[\trialf](x) \right)$.
\State 1. Solve Eq.~(\ref{DPsi}) for $\Psi(x)$ in spectral space, thereby computing the Chebyshev coefficients $\co_k$ for $k=1, \ldots, N$ (but not $k=0$ due to the unknown constant of integration).
\State 2. Using the discrete sine transform (DST), compute $\loadreg(x)$ at the collocation points (see Eq.~(\ref{loadDecomp2})).
\State 3. Transform $\trialf(x)$ to physical space and compute the product $\beta(x) \, \trialf(x)$ at the collocation points.
\State 4.~Transform $\beta(x) \, \trialf(x) + \loadreg(x)$ to spectral space and compute $\Pinv \left( \beta(x) \, \trialf(x) + \loadreg(x) \right)$ by solving BVPs~(\ref{wODE})--(\ref{wBCs}) \& (\ref{uODE})--(\ref{uBCs}) with Algorithm \ref{BVP2Algo}.
\State
\State \MyComment To complete the computation of $\Lopp[\trialf]$, we must calculate $\co_0$.
\State 5. Using Eq.~(\ref{VEq}), compute the function $V(x)$ in spectral space.
\State 6. Using Eq.~(\ref{a0Eq}), calculate the coefficient $\co_0$. 
\State 7. Set $\Lopp[\trialf] = \trialf(x) - \co_0 \, \eta_s(x) - \Pinv \left( \beta(x) \, \trialf(x) + \loadreg(x) \right)$ and return as output (in spectral space).}
\end{algorithmic}
\end{algorithm}

We thus have an algorithm to apply $\Lopp$ with $\BigO{N \log N}$ operations. It is now a simple matter to solve Eq.~(\ref{etaLinEq}) with GMRES (or another Krylov subspace method). For concreteness, suppose we have a routine GMRES($A, b, tol$) that solves the problem $A \vec{x} = \vec{b}$ to tolerance $tol$. Importantly, the input $A$ is permitted to be a {\em function} that computes the matrix-vector product $A\vec{x}$ {without explicitly constructing matrix $A$}. The main algorithm to compute the wing kinematics is then:

\begin{algorithm}[H]
\caption{Main algorithm to solve for the wing kinematics, $\eta(x)$.}
\label{MainAlgo}
\begin{algorithmic}
\State {\bf Inputs:} Stiffness and mass distributions: $\Stiff(x_n), \Mrat(x_n)$ at $N+1$ Chebyshev nodes;
\State {\bf Inputs:} Reduced driving frequency, $\sigma$; 
boundary conditions, $\etaLE, \etaLE'$; desired tolerance, $tol$;
\State 1. Given $\Stiff(x_n)$ and $\Mrat(x_n)$, calculate $\alpha(x_n)$ and $\beta(x_n)$ using Eq.~(\ref{alphabeta}).
\State 2. Set $\etalin(x) = \etaLE + \etaLE' (x+1)$. 
\State 3. Precompute $\eta_s(x)$ by solving BVP (\ref{singODE})--(\ref{singBCs}) with Algorithm \ref{BVP2Algo}.
\State 4. Set $\eta(x)$ = GMRES($\Lopp$, $\etalin(x)$, $tol$), where $\Lopp$ is applied with Algorithm \ref{LAlgo}, and return as output.
\end{algorithmic}
\end{algorithm}
By virtue of the preconditioning, we will find the number of required GMRES iterations to be independent of $N$. Thus, Algorithm \ref{MainAlgo} enables computation of the wing kinematics with an overall $\BigO{N \log N}$ computational cost. 

{\newb  
We note that majority of this computation takes place in spectral space. The only steps that require an $N \log N$ transform to physical space are (i) the multiplications with variable coefficients and (ii) the computation of $\loadreg(x)$ via the sine transform. A banded-operator approach \cite{Olver2013} could be used to perform the variable-coefficient multiplications (to a specified tolerance) while remaining in spectral space. Such a method would therefore avoid several of the $N \log N$ transforms, but not the sine transform associated with $\loadreg(x)$, and thus the overall computational cost would remain $\BigO{N \log N}$.
}

\section{Error analysis and convergence tests}
\label{AccuracySection}

We now discuss the accuracy of our main algorithm. Since it is a spectral method, one might initially expect an exponential convergence rate. However, as we will show in this section, the lack of regularity of the hydrodynamic load, $\loadx(x)$, results in slower 3rd-order convergence. Throughout the section, we will assume that the wing's stiffness and mass distributions, $\Stiff(x)$ and $\Mrat(x)$, are smooth enough so as not to further reduce accuracy.

\subsection{Error analysis}

Recall that the hydrodynamic load has been decomposed into singular and regular components, $\loadx(x) = \co_0 \loadsing(x) + \loadreg(x)$, which are defined in Eq.~(\ref{loadDecomp2}). Through analytical preconditioning, we have removed the singularity in $\loadsing(x)$. However, the nonsingular part, $\loadreg(x)$, also exhibits non-smooth dependence on $x$ as a consequence of the nonlinear transformation $x = \cos \theta$. To see this, we rewrite $\loadreg(x)$ as
\begin{equation}
\label{LoadRegSeries}
\loadreg(x) = 2 \sum_{k=1}^{\infty} \co_k \chi_k(x) \, ,
\end{equation}
where
\begin{equation}
\chi_k(x) = \sin (k \arccos x)
\end{equation}
Consider, for example, the first term $\chi_1(x) = \sin (\arccos x)$. Elementary trigonometry gives  $\chi_1(x) = \sqrt{1-x^2}$, which is {\em H{\"o}lder continuous}  in $x$ of order $1/2$. In fact, all terms possess this same degree of regularity, as can be seen by recasting them in the form
\begin{equation}
\chi_k(x) = \sqrt{1-x^2} \, U_{k-1}(x) \, ,
\end{equation}
where $U_k(x)$ is the {\em second}-kind Chebyshev polynomial of degree $k$. 

It will therefore suffice to consider only $\chi_1(x)$ for error analysis. We perform a Chebyshev expansion
\begin{equation}
\chi_1(x) = \sump{k=0}{\infty} \chico_k T_k(x) \, ,
\end{equation}
where the coefficients $\chico_k$ can be determined exactly by projection,
\begin{equation}
\label{chicoeffs}
\chico_k = \frac{2}{\pi} \int_0^\pi \sin \theta \cos k \theta \, d\theta = 
\begin{cases}
\frac{4}{\pi (1-k^2)} \quad \, \text{if $k$ is even} \\
0 \qquad \qquad \text{if $k$ is odd}
\end{cases}
\end{equation}
These coefficients decay asymptotically like $k^{-2}$, which will be important for determining the method's order of accuracy. 

At this point, we refresh the reader on how the hydrodynamic load enters the computation. Our main method solves Eq.~(\ref{etaLinEq}) iteratively by applying the operator $\Lopp$ to trial functions via Algorithm \ref{LAlgo}. In that algorithm, Step 2 computes $\loadreg[\trialf](x)$ for a given trial function $\trialf(x)$, then Step 4 applies the preconditioner to compute 
\begin{equation}
\label{PinvStuff}
\Pinv \left( \beta(x) \, \trialf(x) + \loadreg(x) \right) \, .
\end{equation}
Since $\Stiff(x)$ and $\Mrat(x)$ are assumed to be smooth functions, $\alpha(x)$ and $\beta(x)$ are likewise smooth. Suppose that $\trialf(x)$ is also sufficiently smooth, so that the only non-smooth term entering Eq.~(\ref{PinvStuff}) is $\loadreg(x)$. Equation (\ref{LoadRegSeries}) expresses $\loadreg(x)$ as a linear combination of the functions $\chi_n(x)$, all of which possess the same degree of regularity. Therefore, to determine the error involved in computing $\Pinv[\loadreg]$, it suffices to consider $\Pinv[\chi_1]$.

To compute $\Pinv[\chi_1]$, we apply Algorithm \ref{BVP2Algo} twice. In the first application of Algorithm \ref{BVP2Algo}, Step 1 computes the antiderivative $\Dinv  \chi_1(x)$, then Step 2 evaluates this antiderivative at $x=1$ to satisfy a boundary condition. {\em It is the combination of these two steps that limits our method's  overall accuracy}. To see this, let us first calculate $\Dinv \chi_1(x)$ exactly by applying Eq.~(\ref{fastint}) directly to Eq.~(\ref{chicoeffs}),
\begin{equation}
\Dinv  \chi_1(x) = \sump{k=0}{\infty} \Chico_k T_k(x) \, ,
\end{equation}
where
\begin{equation}
\Chico_k = \begin{cases}
0 \qquad \qquad \,\,\, \text{if $k \ne 0$ is even} \\
\frac{8}{\pi k^2 (4 - k^2)} \quad \text{if $k$ is odd}
\end{cases}
\end{equation}
Naturally, $\Chico_0$ is a free constant that will be determined by boundary conditions. Importantly, the coefficients of $\Dinv \chi_1(x)$ decay asymptotically like $k^{-4}$.

Now let us examine the errors that arise from computing $\Dinv \chi_1(x)$ numerically. To begin, the pointwise difference between $\chi_1(x)$ and its Chebyshev interpolant $p_N(x)$ results from a combination of {\em truncation error} and {\em aliasing error} \cite{Trefethen2013}, as given by the formula 
\begin{equation}
\label{chierr}
\chi_1(x) - p_N(x) = \sum_{k=N+1}^{\infty} \chico_k T_k(x) - \sump{m=0}{N} \alco_m T_m(x) \, ,
\end{equation}
where
\begin{equation}
\alco_m = \sum_{\ell = 1}^{\infty} \chico_{2 \ell N + m} + \chico_{2 \ell N - m} \, .
\end{equation}
The second sum in Eq.~(\ref{chierr}) is the aliasing error which results from using interpolation to perform {\em approximate} projection onto the Chebyshev basis. Integrating both sides gives
\begin{equation}
\Dinv \chi_1(x) - \Dinv p_N(x) = \sum_{k=N}^{\infty} \tilde{\Chico}_k T_k(x) - \sump{m=0}{N+1} \Alco_m T_m(x) \, ,
\end{equation}
where the $\tilde{\Chico}_k$ coefficients are given by
\begin{equation}
\tilde{\Chico}_N = -\frac{1}{2N} \chico_{N+1} \, , \quad
\tilde{\Chico}_{N+1} = -\frac{1}{2(N+1)}  \chico_{N+2} \, , \quad
\tilde{\Chico}_k = \Chico_k \, \text{ for } k \ge N+2 \, ,
\end{equation}
and the $\Alco_m$ coefficients are determined by Eq.~(\ref{fastint}) as
\begin{equation}
\Alco_m = \frac{1}{2m} \sum_{\ell = 1}^{\infty} \Chico_{2 \ell N + m} - \Chico_{2 \ell N - m} \, .
\end{equation}
Notice that $\Alco_m$ not only depends on $m$, but also on the truncation index, $N$. Furthermore, the decay rate $\Chico_k \sim k^{-4}$ implies that $\Alco_m \sim N^{-4}$. One nice consequence of this scaling is that, by antidifferentiating $p_N(x)$, we have approximated $\Dinv \chi_1(x)$ to the same order of accuracy as if we had interpolated it directly. 

Next, we must evaluate $\Dinv p_N(x)$ at $x=1$ to enforce a boundary condition. Using Eq.~(\ref{fpm1}), the error incurred by this evaluation is
\begin{equation}
\label{EndPointError}
\Dinv \chi_1(1) - \Dinv p_N(1) = \sum_{k=N+1}^{\infty} \Chico_k - \sum_{m=0}^{N} \Alco_m \, .
\end{equation}
Since $\Chico_k \sim k^{-4}$ and $\Alco_m \sim N^{-4}$, both sums scale like $N^{-3}$. This error is carried forward in the subsequent steps (i.e.~the second application of Algorithm \ref{BVP2Algo}), producing a numerical computation of $\Pinv[\chi_1]$ that is {\em third-order accurate}.

A second potential source of error in the main algorithm is the computation of the singular part, $\eta_s$, which is obtained by solving BVP (\ref{singODE})--(\ref{singBCs}) with Algorithm \ref{BVP2Algo}. However, thanks to the analytical preconditioning, the right-hand-side of Eq.~(\ref{singODE}) has Chebyshev coefficients that decay like $k^{-4}$. By similar reasoning as above, this decay rate implies a numerical error of order ${N^{-5}}$. Hence the error in computing $\eta_s$ is subdominant. Similar analysis shows that the Chebyshev coefficients of the solution, $\eta(x)$, decay asymptotically like $k^{-8}$. This scaling not only reveals  the smoothness of our ultimate solution, but also provides a posteriori confirmation that the trial functions $\trialf(x)$ are indeed sufficiently smooth so as not to limit accuracy. Hence, the computation of $\Pinv[\chi_n]$ remains the limiting factor, and {\em our main numerical method for determining the wing kinematics is third-order accurate}.

\subsection{Convergence tests}
\label{ConvTest}

	To verify the third-order accuracy of the method, we now perform a convergence study. Since exact solutions are unavailable (aside from the trivial case of a rigid wing), we estimate the error by comparing numerical solutions of successive resolution. We measure the differences using both the Chebyshev $L^2$-norm and the $L^{\infty}$-norm. The Chebyshev $L^2$-norm is defined as $\norm{u}_2 = \sqrt{ \ip{u,u} }$, with weighted inner-product given by
\begin{equation}
\ip{u,v} = \int_{-1}^{1} \frac{1}{\sqrt{1-x^2}} \, u(x) \, \overline{v(x)} \, dx \, .
\end{equation}	
Due to the weight $(1-x^2)^{-1/2}$, this norm magnifies error near the endpoints, $x=\pm1$, and thanks to Parseval's identity, the norm can be easily calculated in spectral space. Meanwhile, the $L^{\infty}$-norm is defined as $\norm{u}_\infty = \max_{x \in [-1,1]} \abs{u(x)}$ (assuming uniform continuity of $u(x)$ on $[-1,1]$).

\begin{table}
\begin{center}
\caption{Convergence test: The numerical error in the kinematics, $\eta(x)$, as estimated by comparing solutions of successive resolution. The number of collocation points, $N+1$, is quadrupled at each stage. We show both the $L^2$ and $L^{\infty}$ differences, as well as the resulting order of convergence. In this test, we have fixed the parameter values $\Mrat = 1$, $\Stiff = 1$, $\sigma = 1$, and $(\etaLE, \etaLE') = (1,0)$, and have specified a GMRES tolerance of 1E-12. As shown in the second to last column, the number of GMRES iterations required to reach this tolerance is independent of $N$. In the last column, we show measured CPU time of the algorithm.
} 
\vspace{0.3 pc}
\label{convtab}
\begin{tabular}{c l l l l c c}
\hline
\hspace{0.5pc} $N+1$ \hspace{0.5pc} & $L^2$ error & order \hspace{0.7 pc} & $L^{\infty}$ error & order \hspace{0.7 pc} & iterations & CPU time \\
\hline
16		& 3.07E-05   	& --		& 2.40E-05  	& --		& 7	& 0.004	\\
64		& 6.45E-07   	& 2.79	& 5.01E-07  	& 2.79	& 7	& 0.005	\\
256		& 1.08E-08  	& 2.95	& 8.40E-09 	& 2.95	& 7	& 0.006	\\
1,024		& 1.72E-10  	& 2.99	& 1.34E-10 	& 2.99	& 7	& 0.014	\\
4,096		& 2.70E-12  	& 3.00	& 2.09E-12 	& 3.00	& 7	& 0.037	\\
16,384 	& -- 			& --		& -- 			& --		& 7	& 0.180	\\
\hline
\end{tabular}
\end{center}
\end{table}

In Table \ref{convtab}, we show the $L^2$ and $L^{\infty}$ differences between successive numerical solutions of $\eta(x)$, where the number of collocation points, $N+1$, is quadrupled at each stage. We also show the order of accuracy (found by Richardson-extrapolation type calculations). In this test, we have fixed the parameter values $\Mrat = 1$, $\Stiff = 1$, $\sigma = 1$, and $(\etaLE, \etaLE') = (1,0)$. The table confirms that our method converges with the expected third-order accuracy in both norms. The $L^2$-error always exceeds the $L^{\infty}$-error as a result of  $L^2$ giving higher weight to error near the endpoints. 

Surprisingly, the table shows that only 16 collocation points is sufficient to produce nearly 5 digits of accuracy in the computed kinematics. To get 10 digits, one needs roughly 1000 points. In this test, we chose a driving amplitude of  $\etaLE = 1$, so that absolute and relative errors would be nearly the same (only absolute errors are shown in the table). Although $\etaLE = 1$ is not consistent with the small-amplitude assumption, our theory is {\em linear} in amplitude, and so scaling down $\etaLE$ to a more reasonable value simply scales down the solution $\eta(x)$ proportionally. Thus, the values in the table are representative of the relative errors occurring over a range of driving amplitudes.

We also show in the table the number of GMRES iterations required to solve the linear system to a tolerance of 1E-12. The number of iterations is exactly 7 in every case, {\em independent of the size of the system, $N$}. This is a consequence of our preconditioning. As detailed in the next section, solving the unconditioned system (\ref{etaODE})--(\ref{etaBC2}) results in  GMRES either failing to converge at all or requiring the maximum possible number of iterations.

We remark that the parameter values, $\Mrat = 1$, $\Stiff = 1$, $\sigma = 1$, were selected to make the convergence test reasonably stringent. Generally, it is much easier to compute the kinematics when wing deformations are small, which could result from large wing stiffness $\Stiff \gg 1$, low driving frequency $\sigma \ll 1$, a small inertia ratio $\Mrat \ll 1$, or some combination thereof. In the above test, however, all of these parameters are order one, which causes the wing to deform significantly. Despite the large deformations, our method enjoys high accuracy and a modest iteration count, demonstrating its {\em robustness} in parameter space.

The last column of Table \ref{convtab} shows the execution time of our algorithm as implemented in the Julia language \cite{Bezanson2012} on a modern laptop. For most values of $N$, the execution time grows sublinearly, as a significant part of the time is spent on overhead computation. It is only at the last stage, going from $N+1 = 2^{12}$ to $2^{14}$, that the time results are roughly consistent with the predicted $N \log N$ scaling.

{\newb
\subsection{Preconditioner tests}

We now conduct some brief numerical tests to assess the effectiveness of the preconditioner. In these tests we compare solving the preconditioned system (\ref{pcdsEq}) versus the original, unconditioned system (\ref{etaODE})--(\ref{etaBC2}), using GMRES in both cases. Note that the unconditioned system requires several applications of the Chebyshev {\em differentiation} routines and no Chebyshev integration. In Table \ref{condtab}, we report, for each system, the condition number, its order of growth with $N$, and the number of GMRES iterations required to reach a tolerance of 1E-6. In order to obtain the condition number, we explicitly construct the corresponding matrices, accomplished by applying the operator to each of the standard basis vectors (an $\BigO{N^2}$ task that is performed only for the sake of the tests in this section).

\begin{table}
\begin{center}
\caption{{\newb Preconditioner tests. We compare solving the preconditioned system (\ref{pcdsEq}) against solving the original system (\ref{etaODE})--(\ref{etaBC2}), both with GMRES. We report the condition number, its order of growth with $N$, and the number of iterations required to reach a GMRES tolerance of 1E-6. While the condition number of the original system grows like $N^8$, preconditioning reduces that growth rate to $N^3$. More importantly, preconditioning allows GMRES to achieve convergence with a fixed number of iterations, independent of the problem size.
}}
\vspace{0.3 pc}
\label{condtab}
\begin{tabular}{ | c | c l c | c l c | }
\hline
\hspace{0.7pc}
& \multicolumn{3} { c| }{Preconditioned system}
& \multicolumn{3} { c| }{Unconditioned system} \\
\hspace{0.3pc}$N+1$ \hspace{0.3pc} 
& condition number & order & iterations
& condition number & order & iterations\\
\hline
16	& 9.35E+03	& -- 		& 5		&  6.74E+08	& --		& 16	\\
32	& 7.76E+04	& 3.05 	& 5		&  1.90E+11	& 8.14	& --	\\
64	& 6.35E+05	& 3.03 	& 5		&  4.97E+13	& 8.03	& --	\\
128	& 5.14E+06	& 3.02 	& 5		&  1.28E+16	& 8.01	& --	\\
\hline
\end{tabular}
\end{center}
\end{table}
 
The table shows that the condition number grows with increasing resolution whether preconditioning is applied or not. Without preconditioning, the growth rate is $N^8$, consistent with the fourth-order derivative in Eq.~(\ref{etaODE}) and the fact that each Chebyshev differentiation contributes a factor of $N^2$ \cite{Don1995, Greengard1991}. Preconditioning brings the growth rate down to $N^3$, which is a significant improvement yet still rather rapid growth. Of greater importance, though, is how many iterations GMRES requires to converge. Without preconditioning, GMRES only achieves convergence in the lowest resolution case, $N+1=16$, and requires the maximum possible number of iterations to do so. In all other cases, GMRES fails to converge at all. With preconditioning, though, the required number of iterations stays fixed at 5 for all values of $N$.

\begin{figure}
\begin{center}
\includegraphics[width = 0.85 \textwidth]{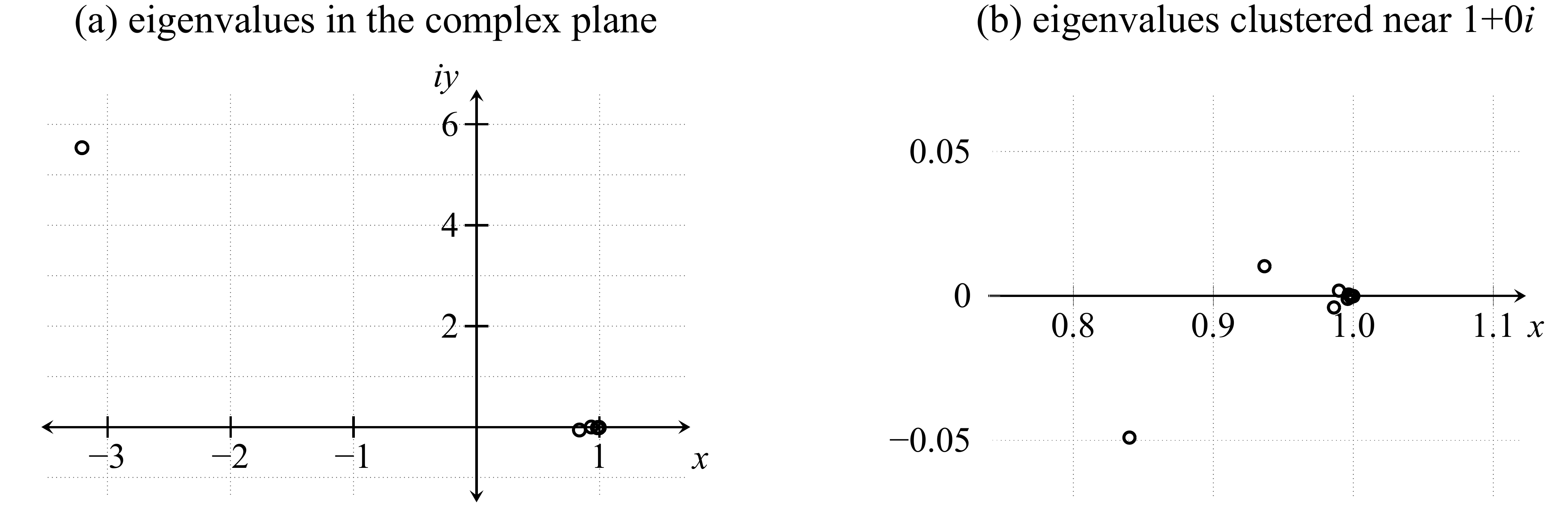}
\caption{{\newb
Spectrum of the preconditioned operator $\Lopp$ defined in Eq.~(\ref{Lopp}). (a) Plotting all of the eigenvalues reveals a cluster near $1+0i$ and a single outlier near $-3.2 + 5.5 i$. (b) Zoom of the cluster point at $1+0i$.
}}
\label{evals}
\end{center}
\end{figure}
 
It may seem counterintuitive that preconditioning results in a {\em fixed} iteration count, even though the condition number still grows with $N$. But this observation simply reflects that the convergence rate of GMRES  depends primarily on the distribution of an operator's {\em eigenvalues}, rather than its singular values \cite{Tref1992, Saad1986, Campbell1996a, Campbell1996b, Kelley1996}. In Fig.~\ref{evals}, we show the spectrum of the preconditioned operator $\Lopp$ from Eq.~(\ref{Lopp}), with 128 collocation points. In Fig.~\ref{evals}(a), {\em all} of the eigenvalues are shown, revealing a cluster near $1+0i$ and a single outlier near $-3.2 + 5.5 i$. Fig.~\ref{evals}(b) shows a zoom of the cluster. These plots indicate that $1+0i$ is the only limit point of the spectrum and that the outliers are few, thus satisfying the criteria for GMRES to converge rapidly \cite{Campbell1996a, Campbell1996b, Kelley1996}.
}

\section{Benchmark: Asymptotic solutions in the stiff-wing limit}
\label{BenchmarkSection}

Section \ref{AccuracySection} establishes the convergence {\em rate} of our method, but does not ensure the method converges to the {\em right answer}. In this section, we derive asymptotic solutions to system (\ref{etaODE})--(\ref{etaBC2}), valid in the stiff-wing limit of $\Stiff \gg 1$. In addition to providing asymptotic results of value in their own right, these solutions offer a way to conclusively validate the numerical method.

To derive the solutions, we consider a nearly rigid wing, $\Stiff \gg 1$, of uniform material properties (i.e.~$\Stiff$ and $\Mrat$ do not depend on $x$). The wing is driven at the leading edge by pure heaving $\etaLE \ne 0$, and we set $\etaLE' =0$. We perform regular perturbation expansions in the small parameter $\Stiff^{-1}$,
\begin{align}
\label{etaPert}
& \eta(x) = \etaLE \left( \eta_0(x) + \Stiff^{-1} \eta_1(x) + \BigO{\Stiff^{-2}} \right) \, , \\
& \loadx(x) = \etaLE \left( \loadx_0 (x)+ \Stiff^{-1} \loadx_1(x) +  \BigO{\Stiff^{-2}} \right) \, .
\end{align}
We have factored out $\etaLE$ here simply for convenience. Inserting these expansions into Eq.~(\ref{etaODE}) and collecting the leading-order terms, $\BigO{\Stiff^1}$, gives
\begin{equation}
\label{eta0ODE}
\frac{8 \pi^2}{3 \sigma^2} \Dx^4 \eta_0(x) = 0 \, ,
\end{equation}
with boundary conditions
\begin{align}
\label{eta0BC1}
& \eta_0(-1) = 1 \, , \quad \eta_0'(-1) = 0 \, , \\
\label{eta0BC2}
& \eta_0''(1) = 0 \, , \qquad \eta_0'''(1) = 0 \, .
\end{align}
The solution to BVP (\ref{eta0ODE})--(\ref{eta0BC2}) is simply $\eta_0(x) = 1$, i.e.~the wing heaves without bending. These kinematics produce a leading-order hydrodynamic load of
\begin{equation}
\label{load0Eq}
\loadx_0(x) = \co_0 \sqrt{ \frac{1-x}{1+x} } + 2 \co_1 \sqrt{1-x^2} \, ,
\end{equation}
where, using Eqs.~(\ref{DPsi}), (\ref{VEq}), and (\ref{a0Eq}), the coefficients are given by
\begin{align}
& \co_0 = -4 \pi j \UFS \theofun(\sigma) \, ,\\
& \co_1 = 4 \pi^2 \, .
\end{align}

The leading-order kinematics, $\eta_0(x) = 1$, describes the pure heaving motion achieved by a perfectly {\em rigid} wing. The associated hydrodynamic load, $\loadx_0(x)$, will influence the next-order kinematics to capture the bending motion of a {\em slightly flexible} wing. Accordingly, at order $\BigO{\Stiff^0}$, Eq.~(\ref{etaODE}) yields an inhomogeneous ODE for $\eta_1(x)$
\begin{equation}
\label{eta1Eq}
\frac{8 \pi^2}{3 \sigma^2} \Dx^4 \eta_1(x) = \loadx_0(x) + 8 \pi^2 \Mrat  \, ,
\end{equation}
with boundary conditions
\begin{align}
\label{eta1BC1}
& \eta_1(-1) = 0 \, , \quad \eta_1'(-1) = 0 \, , \\
\label{eta1BC2}
& \eta_1''(1) = 0 \, , \qquad \eta_1'''(1) = 0 \, .
\end{align}
BVP (\ref{eta1Eq})--(\ref{eta1BC2}) can be solved exactly through repeated antidifferentiation. To expedite this process, we introduce fourth-order antiderivatives of the various terms on the right-hand-side of Eq.~(\ref{eta1Eq}). In particular, we introduce the functions
\begin{align}
& \eta_a(x) = \frac{1}{144} \sqrt{1-x^2} (16 + 39x + 44x^2 + 6x^3) \\
& \qquad  -\frac{3}{144} \arccos x (3 + 12x + 12x^2 + 8x^3) \, , \\
& \eta_b(x) = \frac{1}{720} \sqrt{1-x^2} (16 + 83x^2 + 6x^4) - \frac{15}{720} x \arccos x (3 +4x^2) \, , \\
& \eta_c(x) = \frac{1}{4!} (x-1)^4 \, ,
\end{align}
which satisfy
\begin{align}
\Dx^4 \eta_a(x) =  \singterm \, , \quad
\Dx^4 \eta_b(x) =  \sqrt{ 1-x^2 } \, , \quad
\Dx^4 \eta_c(x) = 1 \, .
\end{align}
Additionally, each of these functions satisfies the free-end boundary condition (\ref{eta1BC2}). The solution, $\eta_1(x)$, is then given by,
\begin{equation}
\frac{8 \pi^2}{3 \sigma^2} \eta_1(x) = \co_0 \eta_a(x) + 2 \co_1 \eta_b(x) + 8 \pi^2 \Mrat \eta_c(x) + A + B(x+1) \, ,
\end{equation}
where $A$ and $B$ are chosen to satisfy the left-end boundary conditions (\ref{eta1BC1}),
\begin{align}
& A = -\frac{5 \pi}{48} \co_0 - \frac{7 \pi}{24} \co_1 - \frac{16 \pi^2}{3} \Mrat \, , \\
& B = \frac{\pi}{4} \co_0 + \frac{5 \pi}{8} \co_1 + \frac{32 \pi^2}{3} \Mrat \, .
\end{align}
We have thus determined the first two terms in perturbation expansion (\ref{etaPert}) for the wing kinematics, with the result being accurate to $\BigO{S^{-2}}$.

\begin{figure}
\begin{center}
\includegraphics[width = 0.85 \textwidth]{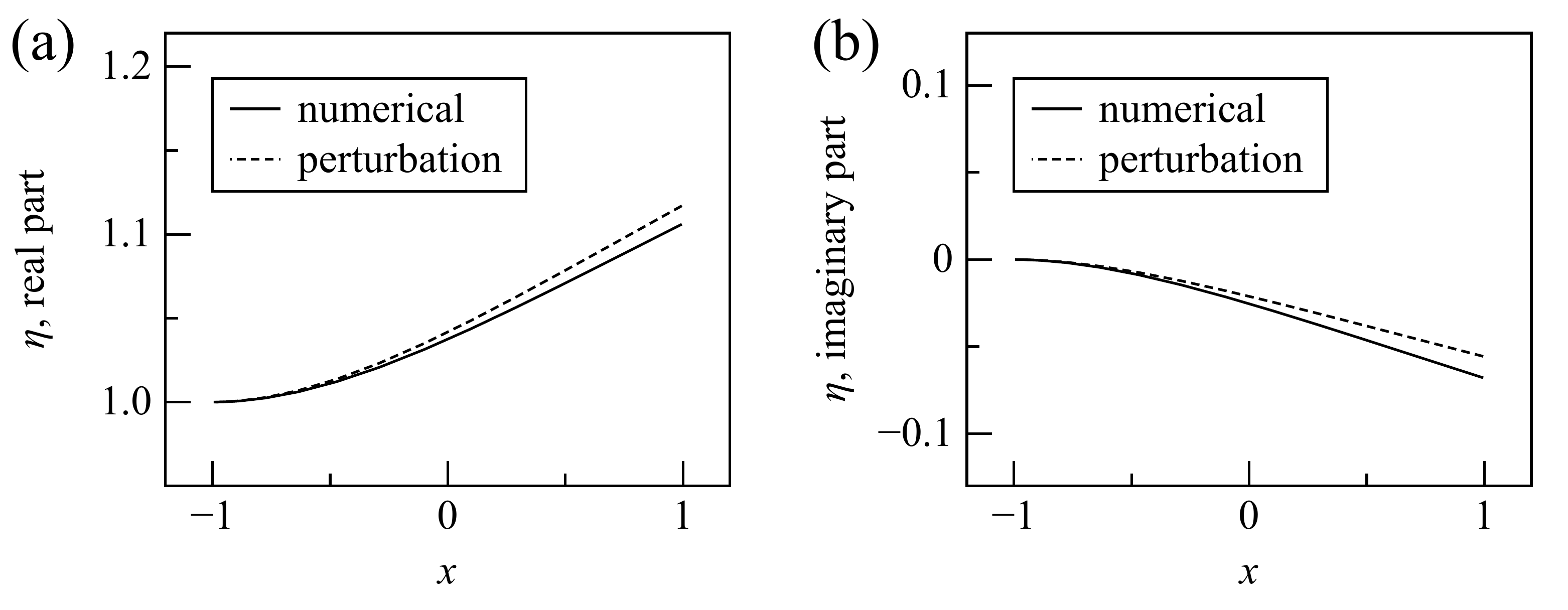}
\caption{ Comparison between the asymptotically and numerically computed kinematics.
The real (a) and imaginary (b) components of $\eta(x)$ as calculated by the perturbation solution (dashed) and the numerical method (solid). The real component shows the wing's bend at the top of the stroke and the imaginary component shows the bend half-way through the upstroke. Here, the following parameters are fixed: $\Stiff = 20$, $\Mrat = 1$, $\sigma = 0.5$, and $(\etaLE, \etaLE') = (1,0)$. Despite $\Stiff$ not being all that large, the asymptotic and numerical solutions agree quite well.
}
\label{bmark}
\end{center}
\end{figure}

	In Fig.~\ref{bmark}, we show these asymptotic solutions as they compare with the numerically computed kinematics. The figure shows both the real and imaginary components of $\eta(x)$, as given by the perturbation expansion (dashed) and by the numerical method (solid). The asymptotics and numerics agree closely in both components of $\eta(x)$. Recall that these real and imaginary components together determine the complete wing kinematics through Eq.~(\ref{heta}). Thus, the entire wing motion can be reconstructed from the two snapshots shown. In this test, we have fixed the parameter values $\Stiff=20$, $\Mrat = 1$, $\sigma = 0.5$, and $\eta_R = 1$. We chose the modest value $\Stiff = 20$ so that the stiff-wing assumption would be satisfied reasonably well, while at the same time, the wing would be allowed to bend perceptibly (i.e.~if $\Stiff$ is too large there is almost no bending at all, which makes for an uninteresting test). As seen in the figure, the wing does bend noticeably, causing displacements of roughly 10\% of the driving amplitude. The numerical and perturbation solutions match remarkably well for $\Stiff$ not being that large. The two differ by about $10^{-2}$ on average, which agrees with the order-of-magnitude estimate of $\BigO{\Stiff^{-2}} \sim {1/400}$.

\begin{table}
\begin{center}
\caption{Convergence of the perturbation and numerical solutions as $\Stiff$ increases. The first panel of the table shows the difference in the real part, $\eta_R(x)$, between the perturbation and numerical solution as calculated in the $L^2$ norm. Here, both the absolute and relative errors show second-order convergence as $\Stiff \to \infty.$ The second panel shows the difference in the imaginary part, $\eta_I(x)$. Since $\eta_I = \BigO{S^{-1}}$, the convergence is second-order in absolute error but first-order in relative error. In this test, we have fixed the parameters $\Mrat = 1$, $\sigma = 0.5$, and $(\etaLE, \etaLE') = (1,0)$. For the numerical solution, we used $N+1 = 256$ and a GMRES tolerance of 1E-8.
} 
\vspace{0.3 pc}
\label{perttab}
\begin{tabular}{ | c | c l c l | c l c l | }
\hline
\hspace{0.7pc}
& \multicolumn{4} { c| }{real part, $\eta_R$}
& \multicolumn{4} { c| }{imaginary part, $\eta_I$} \\
\hspace{0.3pc}$\Stiff$ \hspace{0.3pc} 
& abs err & order \hspace{0.4pc} & rel err & order
& abs err & order \hspace{0.4pc} & rel err & order \\
\hline
50	& 1.67E-03	& -- 		& 9.22E-04	&  --		& 2.14E-03	& --		& 8.63E-02 	& --		\\
100	& 4.04E-04	& 2.04 	& 2.26E-04	&  2.03	& 5.48E-04	& 1.96	& 4.62E-02 	& 0.90	\\
200	& 9.94E-05	& 2.02 	& 5.58E-05	&  2.02	& 1.39E-04	& 1.98	& 2.39E-02 	& 0.95	\\
400	& 2.46E-05 	& 2.01 	& 1.39E-05	&  2.01	& 3.48E-05	& 1.99	& 1.22E-02 	& 0.97	\\
800	& 6.13E-06 	& 2.01 	& 3.46E-06	&  2.00	& 8.74E-06	& 2.00	& 6.14E-03 	& 0.99	\\
\hline
\end{tabular}
\end{center}
\end{table}
 
	To more thoroughly compare the asymptotic and numerical solutions, Table \ref{perttab} demonstrates their convergence as $\Stiff$ increases. We note that the perturbation solution is real-valued at leading order ($\eta_0(x) = 1$) and only becomes complex at order $\Stiff^{-1}$ (from $a_0$ being complex). We therefore aim to verify the convergence of the real and imaginary components of $\eta(x)$ separately. If we did not, errors in $\eta_I(x)$ could potentially go unnoticed as they would be overwhelmed by the larger $\eta_R(x)$. Thus, in Table \ref{perttab} we show the differences in the asymptotic and numerically computed $\eta_R(x)$ and $\eta_I(x)$, as measured in the $L^2$-norm. We show both the absolute and relative errors, as well as the resulting order of convergence. The table confirms second-order convergence of $\eta_R(x)$ as $\Stiff$ increases in both the absolute and relative sense. Meanwhile, the convergence of $\eta_I(x)$ is second-order in absolute error, as is consistent with an $\BigO{\Stiff^{-2}}$ error, but since $\eta_I(x)$ itself is order $\Stiff^{-1}$, the convergence is first-order in relative error. All of these observations are consistent with the expected accuracy of the perturbation solution, thus co-validating the numerical and asymptotic results.

\section{Numerical results and applications}
\label{ResultsSection}

\subsection{Visualization of the kinematics and pressure field}
\label{VisSection}

We now present several numerical examples to demonstrate the utility of our method in understanding flexible-wing propulsion. Recall that our method features the {\em Prandtl acceleration potential}, $\paccel(x,y,t)$, as the primary unknown. Therefore, we first visualize this field is it emerges in a few different cases of flapping. In each case, we use the main algorithm (Algorithm \ref{MainAlgo}) to compute $\eta(x)$, which through Eq.~(\ref{heta}) gives the wing kinematics. This algorithm also calculates the coefficients $a_n$ along the way, and so to determine $\paccel(x,y,t)$, we need only insert the coefficients into Eq.~(\ref{mpole}) and evaluate that multipole expansion in the complex plane $z = x + i y$. In the figures that follow, we will plot $-\paccel(x,y,t)$, since this has the simple interpretation of a {\em dimensionless pressure field}.

\begin{figure}
\begin{center}
\includegraphics[width = 1.0 \textwidth]{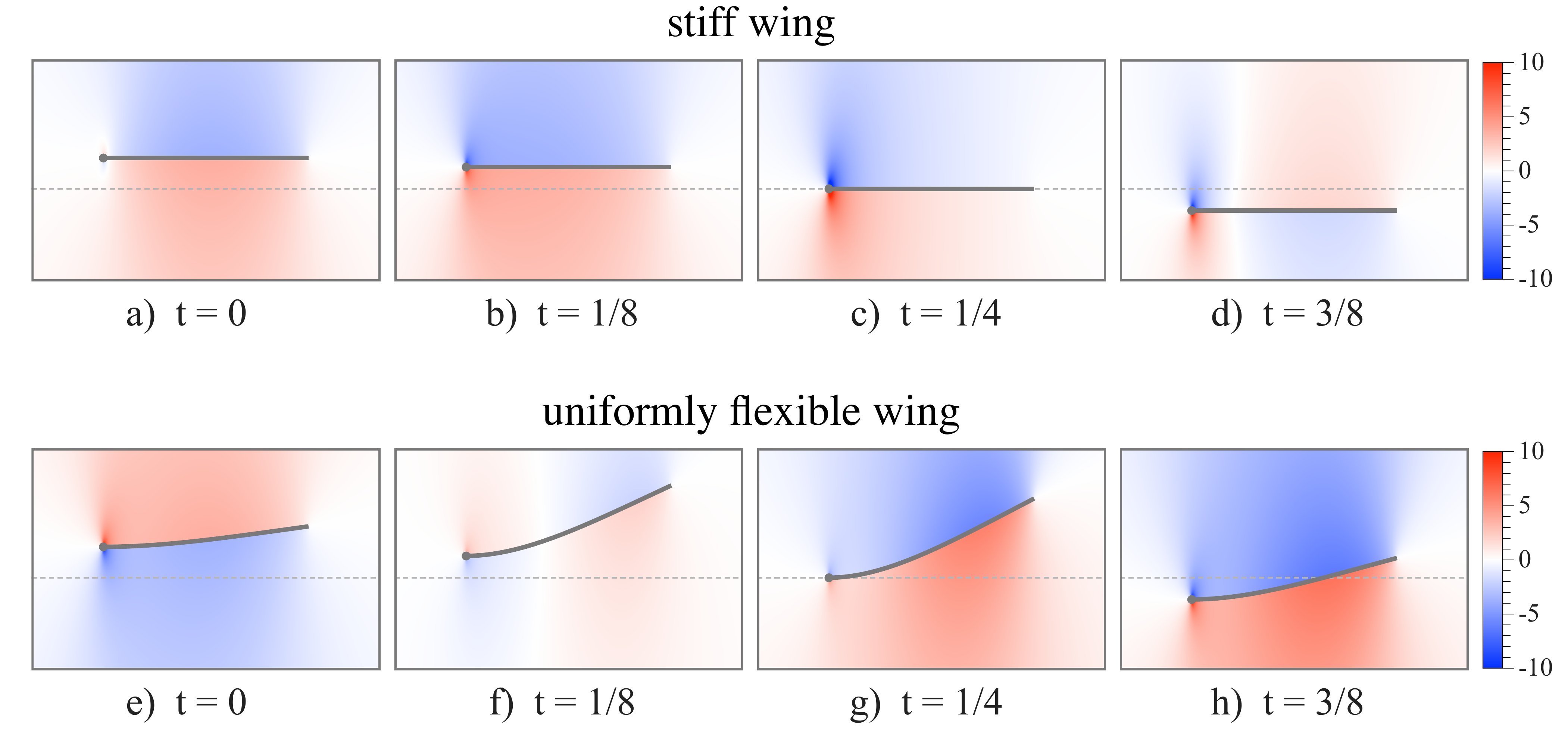}
\caption{ Visualization of the pressure fields surrounding flapping wings. 
Top row (a)--(d): The dimensionless pressure field, $-\paccel(x,y,t)$, surrounding a stiff wing heaved at the leading edge with amplitude $\etaLE = 0.1$ and frequency $\sigma = 1.5$.
Bottom row (e)--(h): The pressure field surrounding a flexible wing of stiffness $\Stiff =15$ driven with the same parameters. The resulting wing deformations modify the pressure field significantly, as can be seen be comparing the top and bottom rows.
In both cases the ratio of solid-to-fluid inertia is $\Mrat =1$. Time is dimensionless, with $t=1$ equal to a flapping period, and we only show the downstroke (the upstroke is symmetric).
}
\label{phi1}
\end{center}
\end{figure}

	Figure \ref{phi1} shows this pressure field as it varies in space and time during a flapping cycle. The figure shows two cases: the top row is a perfectly stiff wing and the bottom row is a uniformly flexible wing with $\Stiff = 15$. Both wings have a solid-to-fluid inertia ratio of $\Mrat = 1$ and are driven with amplitude $\etaLE = 0.1$ and reduced frequency $\sigma = 1.5$. This value of $\sigma$ was chosen because it is approximately the resonant frequency of the flexible wing. The figure shows the wing's downstroke (i.e.~half a flapping period) in evenly spaced time increments. The units are dimensionless, with $t=1$ equal to a full period. The upstroke is symmetric and can be obtained by simply reflecting the plots that are shown about the $x$-axis.

	In the case of a stiff wing (top row), the pressure is higher on the bottom side of the wing throughout most of the downstroke. This arrangement implies positive lift during the downstroke and negative lift during the upstroke, just as would be expected. The strongest pressure variations occur near the wing's leading edge and can be attributed to the singular term in Eq.~(\ref{PhiExp}). These highly localized variations are associated with the so-called leading-edge suction, which contributes to the wing's overall thrust production. In fact, for the case of a rigid wing, the leading-edge suction is the {\em only} source of net thrust, as can be easily deduced from the calculations in \ref{ThrustSection}.
		
	The bottom row of Fig.~\ref{phi1} shows the dynamics of a uniformly flexible wing with dimensionless stiffness $\Stiff = 15$. When driven near resonance, $\sigma = 1.5$, this wing deforms substantially. The relatively deflections of the trailing edge are in fact larger than the amplitude imposed at the leading edge, and these deformations modify the surrounding pressure field significantly. For example, at $t=0$, the pressure field is reversed as compared to the stiff-wing case: high pressure exists above the wing and low pressure below it. The pressure then reverses sign roughly a quarter way through the downstroke ($t=1/8$) so as to produce positive lift throughout the remainder of the downstroke. As in the stiff-wing case, highly localized pressure variations exist near the leading edge. However, the  large-scale pressure variations associated with the flexible wing are somewhat stronger, especially at $t=$ 1/4 and 3/8. Unlike stiff wings, flexible wings can harness these large-scale variations to generate thrust, as leading-edge suction is not the only thrust source.

\begin{figure}
\begin{center}
\includegraphics[width = 1.0 \textwidth]{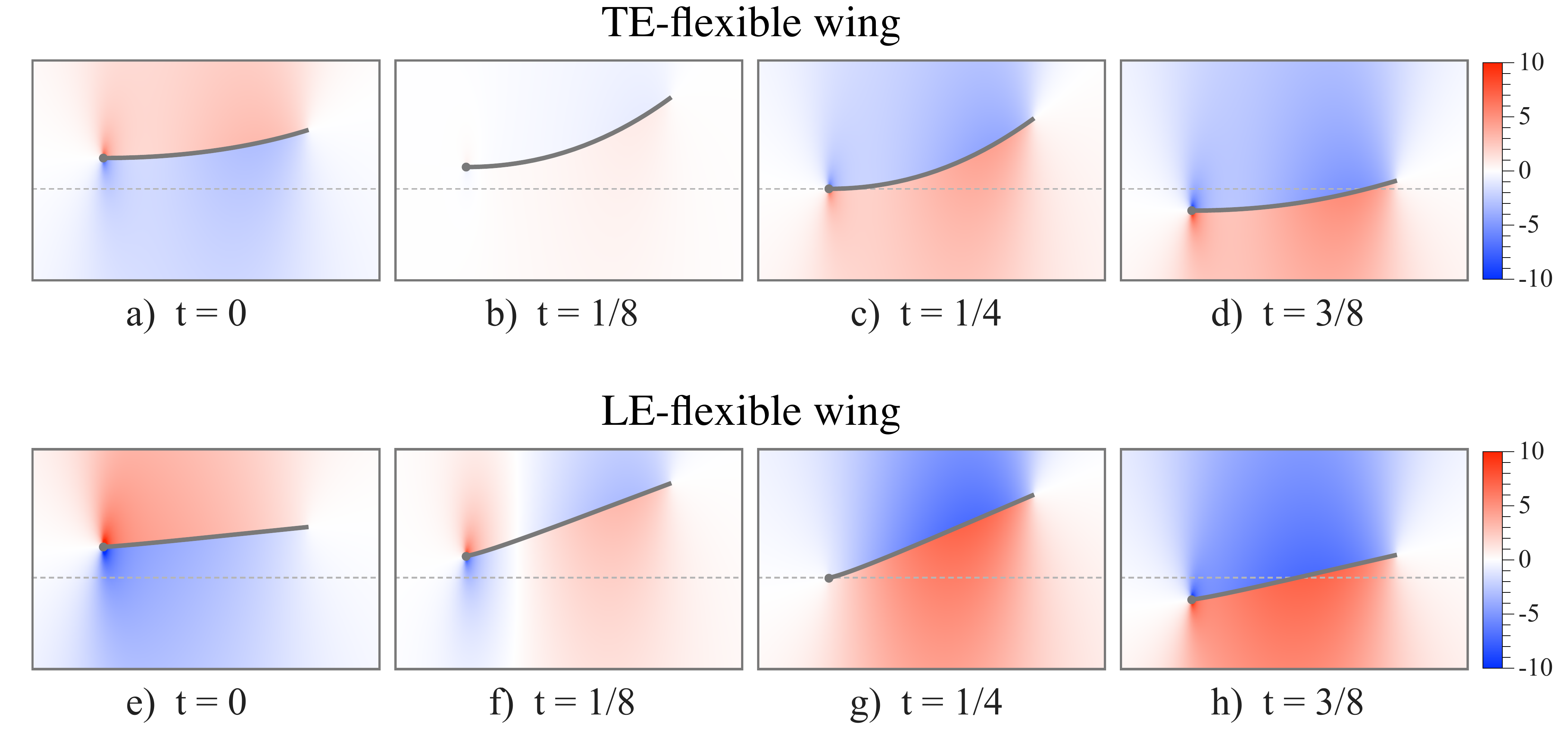}
\caption{ The pressure fields surrounding wings of {\em variable} flexibility.
Top row (a)--(d): A wing that is most flexible at its trailing edge shows a roughly constant bend along the length.
Bottom row (e)--(h): For a wing that is most flexible at the leading edge, the bending is localized, giving the appearance that the wing pivots about the leading edge. The pressure fields surrounding both wings are qualitatively similar at each point in the flapping cycle.
}
\label{phi2}
\end{center}
\end{figure}

	Recall that our numerical method allows for arbitrary distributions of the wing's material properties, $\Stiff$ and $\Mrat$. We therefore consider a few cases of {\em heterogeneous} wings next. Figure \ref{phi2} shows two cases in which $\Mrat =1$ is kept uniform, but the wing stiffness varies, $\Stiff = \Stiff(x)$. The top row shows a wing that is most flexible near its trailing edge (TE flexible), and the bottom row shows a wing that is most flexible near its leading edge (LE flexible). Both are driven with the same parameters as before, $\etaLE = 0.1$ and $\sigma = 1.5$, and we have chosen the distributions of $\Stiff(x)$ to maintain the same resonant frequency.
	
The top row shows the deformations of the TE-flexible wing. Whereas the uniformly flexible wing bends with higher curvature near the leading edge, the TE-flexible wing appears to bend with nearly uniform curvature. The stiffness of this wing increases as one approaches the leading-edge, but the moment arm over which hydrodynamic forces act also increases. These two factors produce the nearly constant curvature seen in the figure. As shown in the bottom row of Fig.~\ref{phi2}, the deformations of the LE-flexible wing are very different. Rather than bending, this wing appears to pivot about its leading edge. In this case, the point of highest flexibility coincides with the longest moment arm, causing the bending to be highly localized near the leading edge. The pressure fields surrounding the TE-flexible and LE-flexible wings are qualitatively similar to each other and to the uniformly flexible wing. In all three cases, high pressure exists above the wing at $t=0$, then the pressure field reverses sign at about $t=1/8$ as the high pressure region migrates below the wing and remains there for the remainder of the downstroke.	

	These plots of the pressure field may seem unfamiliar to some readers, as plots of the vorticity field have become more customary in the recent literature. In addition to their aesthetic appeal, such vorticity plots can elucidate interesting flow physics, such as leading-edge vorticity and rollup of the trailing vortex sheet. However, vorticity is only indirectly related to the hydrodynamic forces that act on and are produced by the wing. It is pressure that couples directly to the wing's force balance to create deformations and, ultimately, generate lift and thrust. We therefore argue that pressure is the most relevant physical quantity for understanding flexible-wing propulsion. In the following sections we will take advantage of this relationship to better understand how certain wings achieve the performance that they do.

\subsection{Thrust, power, and efficiency of heaved wings}
\label{HeaveSection}

	In this section, we will assess the performance of various wings to better understand how material composition influences propulsion. Our primary metric will be the thrust that the wing generates, but we will also consider the input power and the propulsive efficiency. To more easily asses performance across the large parameter space, we use dimensionless thrust and power coefficients, defined as
\begin{equation}
\label{CTCPdefn}
\CT = \frac{ \tavg{\Tdim} } {\frac{1}{2} \pi^3 \rho \freq^2 \amp^2 \chord \width } \, , \qquad
\CP = \frac{ \tavg{\Pdim} } {\frac{1}{2} \pi^3 \rho \freq^2 \amp^2 \Udim \chord \width } \, ,
\end{equation}
where $\Tdim$ and $\Pdim$ are the raw thrust and input power respectively (with units of force and work-per-unit time respectively), and the brackets $\tavg{\cdot}$ indicate an average over one flapping cycle. The {\em propulsive efficiency} is defined as the ratio $\CT/\CP$. A nice consequence of the scales chosen in Eq.~(\ref{CTCPdefn}) is that the thrust and power coefficients computed by the small-amplitude theory are {\em independent of driving amplitude}, thus eliminating one parameter from consideration.

\begin{figure}
\begin{center}
\includegraphics[width = 1.0 \textwidth]{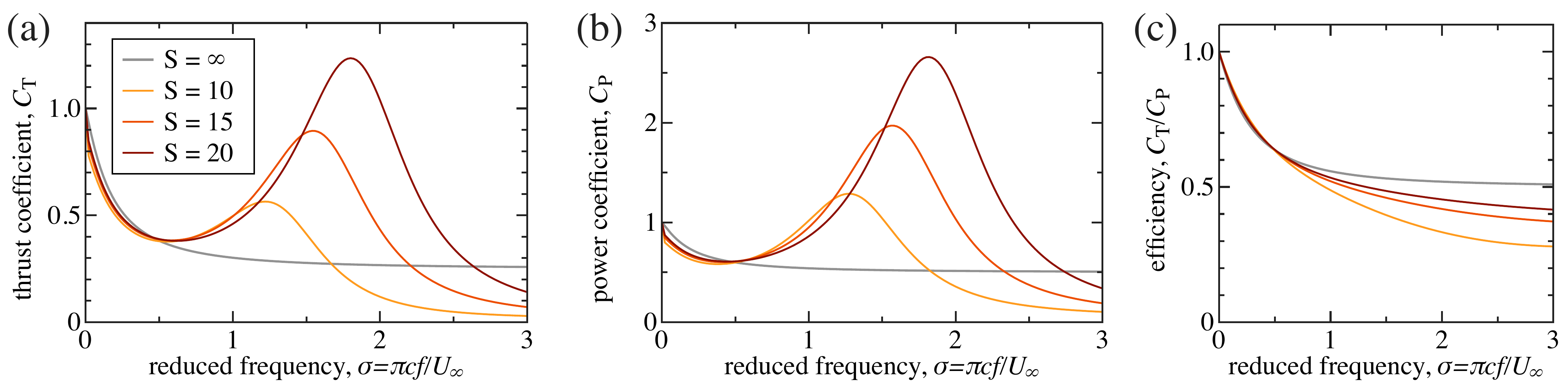}
\caption{ Comparing the performance of uniformly flexible wings. 
(a) The thrust coefficient, $\CT$, versus reduced frequency, $\sigma$, for wings of various stiffness values (see legend). In all cases $\Mrat = 1$, and the wing is driven by pure heaving at the leading edge.
(b) The power coefficient, $\CP$, versus frequency for the same set of wings, showing qualitatively similar trends.
(c) The propulsive efficiency, $\CT/\CP$, for these same wings. The peaks in $\CT$ and $\CP$ offset, causing the efficiency to decrease monotonically with driving frequency.
}
\label{thruni}
\end{center}
\end{figure}

	Once the wing kinematics our computed, it is relatively straightforward to calculate the thrust and power that result. We provide those details in \ref{ThrustSection}, so that in this section we can focus on comparing the performance of different wings. To begin with a simple case, we first compare several wings of {\em uniform} flexibility that are driven by heaving at the leading edge. Figure~\ref{thruni} shows the thrust, power, and efficiency achieved by these wings over a range of driving frequencies. In all cases, $\Mrat = 1$ is fixed and the driving is given by $(\etaLE, \etaLE') = (0.1, 0)$, although, again, the plots are insensitive to the value $\etaLE = 0.1$ as a result of our nondimensionalization.
				
	First, the gray curve in Fig.~\ref{thruni}(a) represents a perfectly stiff wing, $\Stiff = \infty$, which serves as a baseline. Here, the thrust coefficient decreases monotonically and approaches the value $\CT = 1/4$ as $\sigma \to \infty$. The value $\CT=1$ at $\sigma = 0$ agrees with well-known quasi-steady solutions (not restricted to small-amplitude) \cite{Childress1981, Saffman1992}, thus validating our thrust calculation. The three colored curves in Fig.~\ref{thruni}(a) represent wings of various uniform stiffnesses, $\Stiff =$ 10, 15, and 20. In each case, there exists a resonant frequency that maximizes thrust production. As the stiffness of the wing increases, resonance occurs at a higher frequency and the peak-value of $\CT$ is also greater. If driven well beyond resonance, each of these wings  produces {\em less} thrust than the perfectly stiff wing. All of these observations are qualitatively similar to the case in which flexibility is highly localized through a torsional spring - a case for which explicit solutions are available \cite{Moore2014}.
	
	Figure \ref{thruni}(b) shows the input power required to drive these same wings. The power coefficient shows qualitatively similar trends as $\CT$, with a peak value occurring at a resonant frequency. As seen in \ref{thruni}(c), these peaks cancel in the ratio $\CT/\CP$, so that the propulsive efficiency decreases monotonically. In the figure, all of the flexible wings exhibit lower efficiency than the rigid wing. However, this result can be sensitive to the inertia ratio, $\Mrat$. Here, we have set $\Mrat = 1$, but if $\Mrat=0$ for example, flexible wings can be more efficient than their rigid counterparts \cite{Moore2014}. 

	We remark that at much higher driving frequencies, flexible wings exhibit additional peaks in thrust and power that correspond to higher bending modes \cite{Alben2008, Moore2015}. However, the bending amplitudes predicted by our theory in this regime are usually too large to satisfy the small-amplitude assumption \cite{Moore2015}. Thus, while the small-amplitude theory captures the existence of these higher harmonics, it does not accurately predict the corresponding hydrodynamic forces. We therefore focus on the first bending mode only and moderate reduced frequencies (typically $\sigma < 5$).

\begin{figure}
\begin{center}
\includegraphics[width = 0.9 \textwidth]{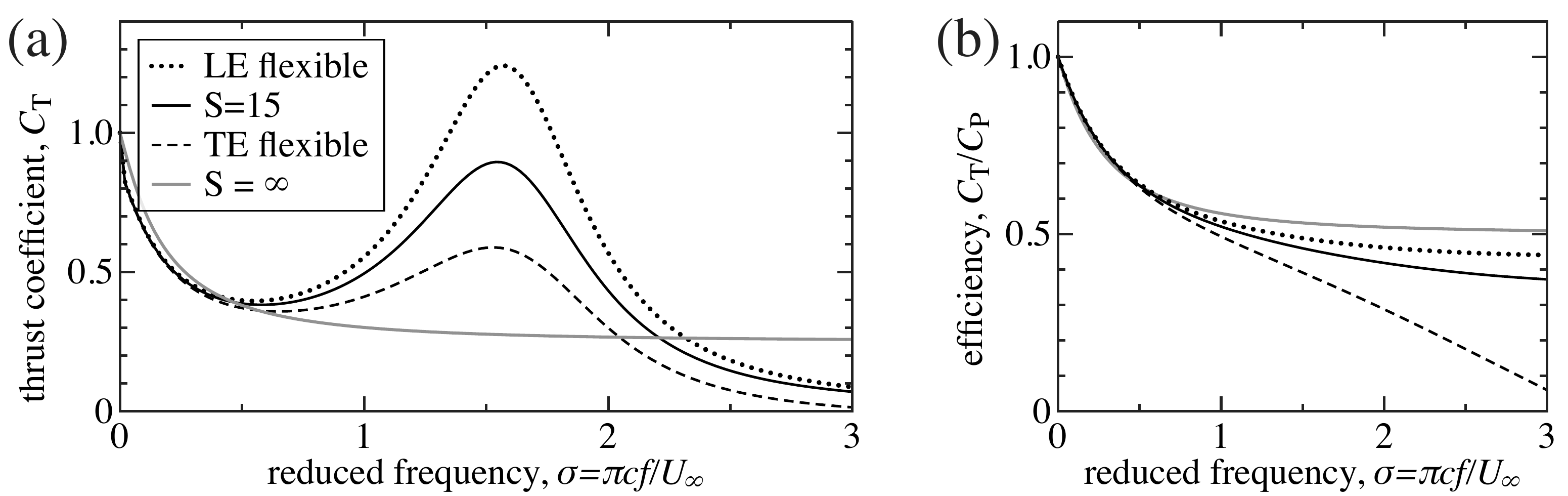}
\caption{ Comparing the performance of heterogeneous wings.
(a) The thrust coefficient versus driving frequency for the uniformly flexible wing with $\Stiff = 15$ (solid black curve), the TE-flexible wing (dashed), the LE-flexible wing (dotted), and the stiff wing (solid gray), all with $\Mrat = 1$. The LE-flexible wing generates significantly greater peak-thrust than all other wings.
(b) The propulsive efficiency versus driving frequency for each of the same wings. Although, flexible wings generate more thrust, the rigid wing achieves the greatest efficiency.
}
\label{thrvar}
\end{center}
\end{figure}
 
	Next, we consider a few cases of {\em heterogeneous} material composition. To begin, we consider the uniform case $\Stiff = 15$, and modify this wing to make either the trailing edge (TE) or the leading edge (LE) more flexible while maintaining the same resonant frequency, $\sigma = 1.5$ (we keep $\Mrat = 1$ fixed too). These are the same wings from Fig.~\ref{phi2}, in which we showed the kinematics and pressure fields. In Fig.~\ref{thrvar}(a), we show the thrust produced by each of these wings over a range of driving frequencies. The figure confirms that all three wings resonate at the same frequency. However, they do not produce the same thrust. The TE-flexible wing generates less thrust than the uniform case. This wing is the one seen to bend with roughly constant curvature in Fig.~\ref{phi2}. On the other hand, The LE-flexible wing generates significantly greater thrust than the other two. The peak-thrust is roughly 40\% greater than in the uniform case and more than double that of the TE-flexible case. Evidently, the pivoting motion exhibited by the LE-flexible wing in Fig.~\ref{phi2} provides a significant hydrodynamic advantage.
		
	The differences in thrust production can be reconciled with the pressure-field plots in a somewhat unexpected way. Thrust arises from the difference in pressure across the top and bottom wing surfaces that is directed along the horizontal. Hence, thrust depends on both the pressure jump and the slope of the wing surface, as expressed in Eq.~(\ref{TpEq}). As seen in Figs.~\ref{phi1} and \ref{phi2}, the pressure fields surrounding each of the flexible wings are all qualitatively similar. Differences on the order of 40--100\% are not detectable. However, the wing-surface slopes differ substantially. Whereas the uniform and the TE-flexible wings are flat near the leading-edge, the LE-flexible wing deflects along its entire length. This allows more of the pressure difference to be directed horizontally so as to generate greater thrust.

	In Fig.~{\ref{thrvar}(b), we show the propulsive efficiency achieved by each of these same wings (we do not show $\CP$ since it can be inferred). Once again, the peaks in thrust are offset by peaks in power so that the efficiency decays monotonically. The LE-flexible wing, in addition to generating the greatest thrust, is also the most efficient of the flexible wings. Here, none of the flexible wings propel as efficiently as the rigid wing, though once again this result can be sensitive to the value of $\Mrat$.

\subsection{Thrust, power, and efficiency of pitched wings}
\label{PitchSection}

 	To demonstrate the versatility of our numerical method, we now show how the above results change when the wings are driven by {\em pitching} instead of heaving. We first consider the same {\em uniformly} flexible wings from Fig.~\ref{thruni}, but now prescribe the driving as $(\etaLE, \etaLE') = (0, 0.1)$. Figure \ref{thrunipitch} shows the thrust, power, and efficiency achieved in each case. As seen in Fig.~\ref{thrunipitch}(a), each wing produces {\em negative} thrust, i.e.~drag, when driven at a low enough frequency. This behavior was not observed for any of the heaved wings. At higher frequencies, each of the pitched wings generates positive thrust and exhibits a resonant peak in $\CT$. As in the heaved case, higher wing stiffness increases the resonant frequency as well as the peak-value of $\CT$. Although the trends are similar, the wings produce less thrust overall when driven by pitching compared to heaving (typically about 50\% less thrust).

\begin{figure}
\begin{center}
\includegraphics[width = 1.0 \textwidth]{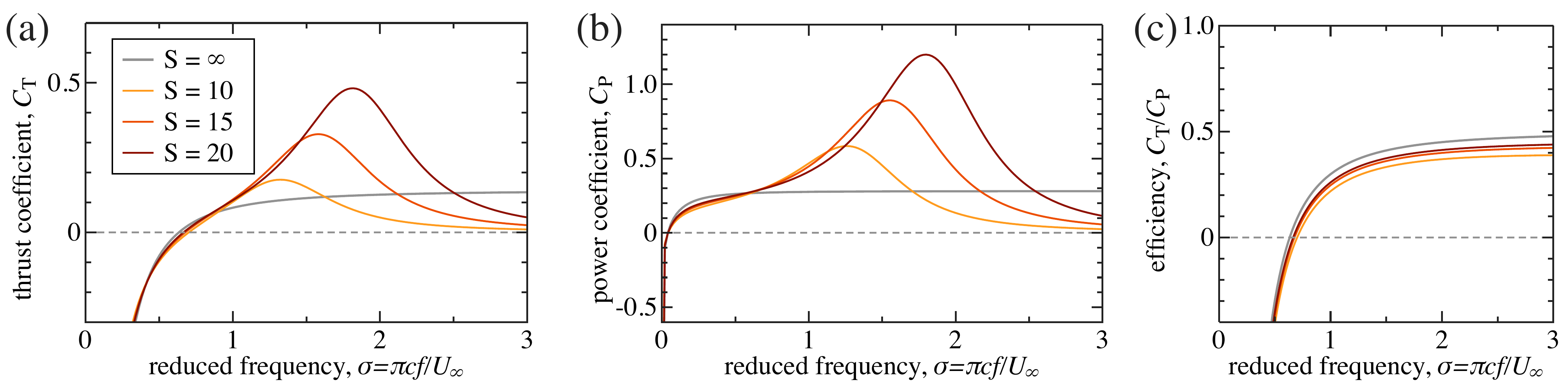}
\caption{ Comparing the performance of uniformly flexible wings driven by {\em pitching}.
(a) The thrust coefficient versus driving frequency for wings of various stiffness values (see legend), all with $\Mrat =1$. When driven by pitching, these wings produce negative thrust at small driving frequency. At higher $\sigma$, the wings generate positive thrust which peaks at resonance.
(b) The power coefficient for the same set of wings showing similar trends as thrust.
(c) The propulsive efficiency for these same wings. As before, efficiency is monotone, but the efficiency increases with driving frequency for pitched wings.
}
\label{thrunipitch}
\end{center}
\end{figure}
 	
	As seen in Fig.~\ref{thrunipitch}(b), the power requirements of the pitched wings also follow similar trends. In each case, $\CP$ obtains a maximum at a resonant frequency then drops below the rigid-wing value at higher frequencies. Once again, the overall values of $\CP$ are significantly smaller for pitched wings than for heaved wings. As before, the peaks in $\CT$ and $\CP$  offset to produce an efficiency with no peaks as seen in Fig.~\ref{thrunipitch}(c). For the pitched wings, though, the efficiency {\em increases} monotonically with $\sigma$. Unlike heaving, pitching becomes more efficient at higher driving frequencies.
 
 	We next ask how {\em heterogeneous} wings perform when driven by pitching. We use the same TE-flexible and LE-flexible wings as before, both of which are modifications of the reference case $\Stiff = 15$. Figure \ref{thrvarpitch}(a) shows the thrust produced by each of these wings, revealing negative thrust at small frequencies and positive thrust at higher frequencies. All three resonate at nearly the same $\sigma$, and, intriguingly, the LE-flexible wing once again outperforms the other two by a healthy margin. Thus, the superior performance of LE-flexibility appears to be insensitive to the actuation strategy.

\begin{figure}
\begin{center}
\includegraphics[width = 0.9 \textwidth]{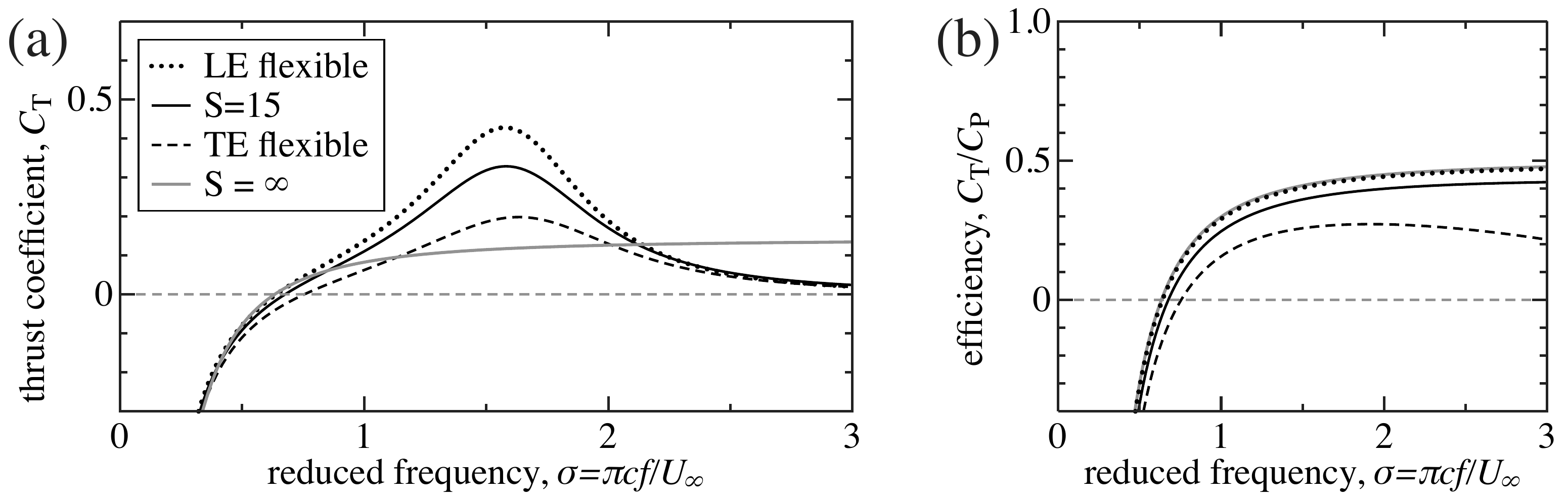}
\caption{ Comparing the performance of heterogeneous wings driven by pitching. 
(a) Thrust coefficient versus driving frequency for the uniformly flexible wing (solid black), the TE-flexible wing (dashed), the LE-flexible wing (dotted), and the stiff wing (solid gray), all with $\Mrat = 1$. For pitched wings, the LE-flexible wing remains the best performer.
(b) Propulsive efficiency for these same wings. Although, the stiff wing is still the most efficient, the LE-flexible wing is very close.
}
\label{thrvarpitch}
\end{center}
\end{figure}
 	
	In Fig.~\ref{thrvarpitch}(b), we show the propulsive efficiency achieved by each of the same wings. The LE-flexible wing outperforms the other two flexible wings in this metric too, although none of the flexible wings propel as efficiently as the rigid wing. Unlike the heaved case, though, the efficiency of the LE-flexible wing nearly matches that of the rigid wing.

\subsection{Comments on the optimization of Moore (2015)}

Here, we have considered a variety of stiffness distributions and found that localizing flexibility near the driving point (i.e.~the leading edge) can enhance thrust production significantly for both heaved and pitched wings. This observation naturally raise the question of which material distribution is optimal for performance. That question is not easily answered, as it clearly depends on factors such as the actuation strategy and the performance metric used. Nonetheless, the question was addressed by Moore (2015) for the case of optimizing the flexibility distribution of heaved wings for thrust production \cite{Moore2015}. It was found that localizing all of the wing's flexibility at the driving point (through the use of a torsional spring) globally optimizes thrust, as is consistent with the observations in this study. 
	
	Since those results have been reported already, we do not rehash them here, but we do provide a few additional comments on the optimization procedure. In Moore (2015), the optimization was performed using the Broyden-Fletcher-Goldfarb-Shanno (BFGS) algorithm, which is a quasi-Newton method \cite{Wright99}. Since then, we have experimented with other optimization routines, including the Nelder Mead simplex method \cite{NelderMead1965, Gao2012}. We have found that, for our problem, Nelder Mead performs the optimization about 3 times faster than the BFGS implementation used. For fixed $\Mrat$ and $\sigma$, Nelder Mead typically requires 200-600 iterations to optimize $\Stiff(x)$ over the space of cubic polynomials (a four-dimensional parameter space). Moore (2015) performed this optimization over a range of frequencies, $0 \le \sigma \le 7$, with a grid spacing of $d\sigma = 0.05$, implying somewhere in the neighborhood of 5000 calls to the PDE solver. This number of calls is only made feasible by having a fast PDE solver.

\subsection{Simultaneous variation of the wing's stiffness and mass}

	Now, to really demonstrate the capability of our method to search parameter space, we examine the same performance metrics with the wing stiffness $\Stiff$ and inertia ratio $\Mrat$ varied {\em simultaneously}. In Fig.~\ref{SRplots} we show the thrust coefficient $\CT$ (top row) and efficiency $\CT/\CP$ (bottom row) obtained by wings spanning the 2D parameter space of $0 \le \Stiff \le 40$ and $0 \le \Mrat \le 4$. The left column shows wings driven by heaving $(\etaLE, \etaLE') = (0.1, 0)$ and the right by pitching $(\etaLE, \etaLE') = (0, 0.1)$. In all cases, the driving frequency is fixed at $\sigma = 1.5$ (the resonant frequency of wings used in previous sections).

\begin{figure}
\begin{center}
\includegraphics[width = 0.6 \textwidth]{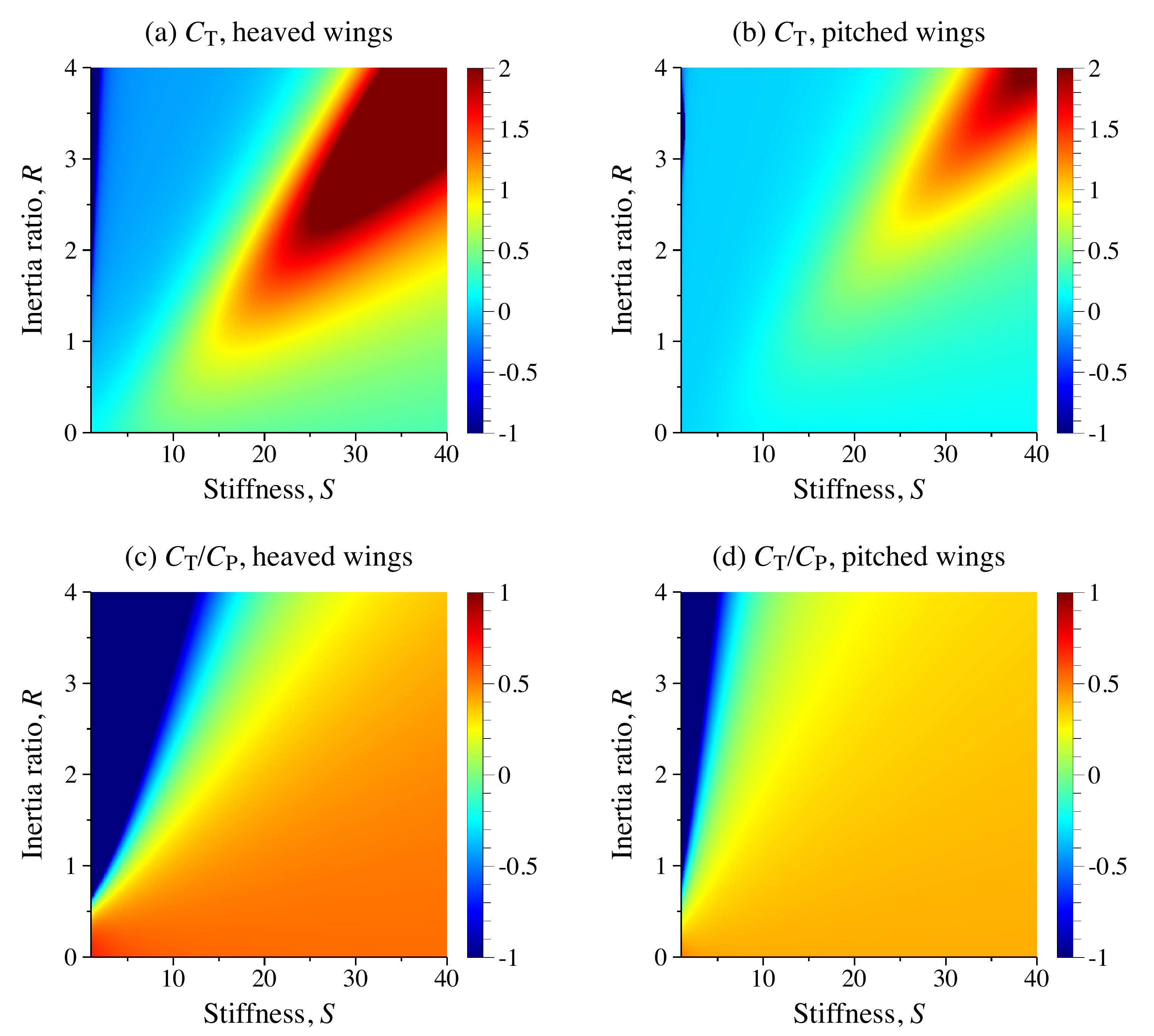}
\caption{Thrust and efficiency over the 2D parameter space of $\Stiff$ and $\Mrat$. (a)--(b) The top row shows the thrust coefficient of (a) heaved and (b) pitched wings, both driven at frequency of $\sigma = 1.5$. (c)--(d) The bottom row shows the efficiency, $\CT/\CP$ for the same cases. At fixed driving frequency, thrust favors stiff, heavy wings, while efficiency favors stiff, light wings.
}
\label{SRplots}
\end{center}
\end{figure}
 
As seen in Figs.~\ref{SRplots}(a)-(b), thrust favors wings of higher stiffness and inertia ratio for both heaving and pitching. If $\Mrat$ is fixed, corresponding to traversing the plot along a horizontal line, there exists an $\Stiff$ value that maximizes $\CT$, as is consistent with Sections \ref{HeaveSection} and \ref{PitchSection}. Conversely, if one fixes $\Stiff$ (a vertical line), we see that an optimal $\Mrat$ exists. These observations apply to both heaved and pitched wings, although the thrust produced by pitching is generally smaller. Interestingly, there is a small region in the parameter space where the wings produce negative thrust. This region corresponds to heavy, highly flexible wings.
 
On the other hand, Figs.~\ref{SRplots}(c)-(d) show that efficiency $\CT/\CP$ favors wings of high stiffness and {\em low} inertia ratio. This figure shows no optimality with respect to fixing either $\Stiff$ or $\Mrat$, which is again consistent with the previous sections. To summarize, at a fixed driving frequency, thrust favors stiffer, heavier wings, while efficiency favors stiffer, lighter wings.

This bird's-eye view of performance over the 2D parameter is only made possible by having a fast PDE solver. As illustration, Fig.~\ref{SRplots} uses an 80 x 80 grid over $\Stiff$ and $\Mrat$ for both heaved and pitched wings, requiring a total 12,800 calls to the PDE solver. Using our method with $64$ collocation points, this requires just under a minute of computational time on a modern laptop.

	We close this section with a comment on the accuracy of the small-amplitude theory in predicting thrust, power, and efficiency in practical applications. Moored et al.~(2017) recently compared small-amplitude results against a vortex-sheet tracking method for the case of a rigid wing driven by pitching at the leading edge \cite{Moored2017}. It was found that the small-amplitude theory accurately predicts thrust over a range of driving amplitudes and frequencies. However, discrepancies were found in the power and, consequently, the efficiency, as these quantities showed much greater sensitivity to the Strouhal number, $\St = \freq \amp / \Udim$. Consequently, Moored et al.~(2017) developed a new class of scaling laws to rationalize the observations \cite{Moored2017}. These findings are also consistent with the observation of Moore (2014), that the small-amplitude theory captures experimentally observed peaks in thrust production but not in efficiency \cite{Moore2014}. 
	
	For these reasons, we payed closest attention to how the small-amplitude model predicts thrust and, in fact, optimized this quantity in Moore (2015) \cite{Moore2015}. The current paper provides additional results on power and efficiency, as it is valuable to know what the small-amplitude theory predicts for these quantities, even if the predictions have a limited range of validity. A promising future direction is to extend the theory to the first correction in $\St$ so as to compute power and efficiency more accurately. This extension would involve solving a Poisson problem for the first-order correction to $\paccel$ and might capture previously reported optimality with respect to the Strouhal number \cite{Triant1993, Anderson1998, Taylor2003, Eloy2012}.

\section{Discussion}
\label{SectionDiscussion}

	In this paper, we developed a highly efficient Chebyshev method to simulate flapping propulsion by wings of arbitrary material distributions. The method exploits the fact that the collective influence of the hydrodynamics can be represented by a nonlocal operator that acts on the 1D wing kinematics and that can be rapidly evaluated with $\BigO{N \log N}$ operations. We used an iterative method to solve for the resulting wing kinematics, after applying preconditioning to the continuous form of the system and analytically removing a flow singularity. Analysis of higher-order singularities reveals the method to be third-order accurate, as confirmed by a convergence test, and we used newly derived asymptotic solutions to provide validation.

	The main advantage of the method is that it permits rapid exploration of the very large parameter space of all possible material distributions. This parameter space is, in principle, infinite dimensional as it encompasses all distributions of flexibility and wing density, represented by $\Stiff(x)$ and $\Mrat(x)$. In this paper, we have examined the performance a multitude of wings in this space, and Moore (2015) used an earlier version of the method to optimize a wing's flexibility distribution for thrust production \cite{Moore2015}.
	
	Performing such an optimization using DNS or a vortex-sheet tracking method would be much more expensive, likely prohibitively so. On the other hand, the $N$-values required by the optimization might not show a large difference between our $\BigO{N \log N}$ method and the $\BigO{N^2}$ method of Alben (2008) \cite{Alben2008}. However, the speed of our method would stand out in applications that involve {\em many} flapping bodies. Although we have not discussed such applications here, generalizations of the small-amplitude analysis have been made to treat multiple flappers \cite{Wu1975}. One such version was recently used to examine self-organized schooling behavior in the case of {\em aligned} swimming bodies \cite{Fang2016}. Extending the analysis to treat multiple bodies of arbitrary position and alignment is a promising future direction.
	
	A second promising direction is to extend the small-amplitude analysis into three dimensions, so as to handle wings/fins that deform along both their chord and their {\em span} \cite{DeLaurier1999, DeLaurier2003, Moored2011a, Fish2016}. This extension might be achieved by coupling cross-sectional domains through a circulation relation, much like is done in classical lifting-line theory \cite{Phillips2000, AlbenBrenner2008}. While lifting-line theory relies on {\em quasi-steady} hydrodynamic relationships, the small-amplitude theory does not constrain the driving frequency $\sigma$ to be small and therefore offers the ability to capture a range of different {\em dynamic} states.

\section*{Acknowledgements} \noindent The author would like to thank Keith Moored, Saverio Spagnolie, Mark Sussman, and Bryan Quaife for helpful discussions.

\appendix
\include{Nomenclature}

\section{Calculation of thrust and power}
\label{ThrustSection}

In this section we discuss how to calculate the thrust and power once the wing kinematics are known. In Section \ref{NondimSection}, we moved to dimensionless variables by scaling length on $\chord/2$ and time on $1/\freq$. These scales produce a characteristic force-per-unit-span of $\rho \freq^2 \chord^3 / 8$ and power-per-unit-span of $\rho \freq^3 \chord^4 / 16$. Consequently, the dimensionless thrust and power (indicated by a tilde) are related to the raw thrust and power via
\begin{align}
& \Tnd = \frac{8 \Tdim}{\rho \freq^2 \chord^3 \width} \\
& \Pnd = \frac{16 \Pdim}{\rho \freq^3 \chord^4 \width}
\end{align}

\noindent
{\bf Thrust:} To calculate thrust, we decompose it as $\Tnd = \Tnd_s + \Tnd_p$, where $\Tnd_s$ is the so-called leading-edge section and  $\Tnd_p$ arises from the pressure difference across the wing. The latter can be calculated as
\begin{equation}
\label{TpEq}
\Tnd_p = \int_{-1}^{1} \load_R \,\, \pd{h_R}{x} \, dx \, .
\end{equation}
Recalling  $\load(x,t) = \loadx(x) \ejt$ and $h(x,t) = \eta(x) \ejt$ and averaging over one flapping cycle (indicated by brackets) yields
\begin{equation}
\tavg{\Tnd_p} = 
 \frac{1}{2} \int_{-1}^{1} \loadx_R \td{\eta_R}{x} + \loadx_I \td{\eta_I}{x} \, dx \, ,
\end{equation}
where the subscripts $R$ and $I$ indicate the real and imaginary part with respect to $j$.
We compute this integral through Chebyshev-Gauss quadrature, accomplished by changing the integration variable to $\theta = \arccos x$ and using the midpoint rule with an evenly-spaced $\theta$ grid. With the change-of-variables, the integral becomes
\begin{equation}
\tavg{\Tnd_p} = \frac{1}{2} \int_0^{\pi} 
\left(  \loadx_R \td{\eta_R}{x} +  \loadx_I \td{\eta_I}{x} \right) \sin \theta \, d\theta \, ,
\end{equation}
Recall that $\loadx$ is singular at $x=-1$ or $\theta = \pi$, as seen in Eq.~(\ref{loadxEq}). To preserve accuracy of the quadrature, we use the identity $\tan (\theta/2) \sin \theta = 1 - \cos \theta$ to get
\begin{equation}
\loadx \, \sin \theta = \co_0 (1-\cos \theta) + 2 \sin \theta \sum_{n=0}^{\infty} \co_n \sin n \theta \, .
\end{equation}
The singularity has been removed from the first term on the right.

Meanwhile, the leading-edge suction can by calculated by contour integration \cite{Wu1961, Moore2014}, giving
\begin{equation}
\Tnd_s = \frac{\pi}{2 \UFS^2} \left( ( \co_0 \ejt )_R \right)^2 \, ,
\end{equation}
Taking the time average gives
\begin{equation}
\tavg{\Tnd_s} = \frac{\pi}{4 \UFS^2} \abs{\co_0}^2 \, .
\end{equation}

\noindent
{\bf Power:} The power is defined as
\begin{equation}
\Pnd = -\int_{-1}^{1} \load_R \left( \pd{h}{t} \right)_R dx
\end{equation}
Taking the time average and transforming to a $\theta$ integral gives
\begin{equation}
\tavg{\Pnd} = \pi \int_{0}^{\pi} \left( \loadx_R \eta_I - \loadx_I \eta_R \right) \, \sin \theta \, d\theta
\end{equation}
\vsp{3}

\noindent
{\bf Thrust and power coefficients:} Now to convert to the thrust and power coefficients defined in Eq.~(\ref{CTCPdefn}), we use
\begin{equation}
\CT = \tavg{\Tnd}/(4 \pi^3 \etaref^2) \, , \qquad
\CP = \tavg{\Pnd}/(4 \pi^3 \UFS \etaref^2 ) \, ,
\end{equation}
where $\etaref = \amp / \chord$ is the dimensionless reference amplitude. In the case of a heaved wing, $\amp$ is simply the peak-to-peak driving amplitude and so $\etaref$ is its dimensionless counterpart. In the case of a pitched wing, though, the heaving amplitude is zero and so $\etaref$ must be defined another way. We therefore define
\begin{equation}
\etaref = \max \left\{ \abs{\etaLE},  \abs{\etaLE + 2 \etaLE'} \right\}
\end{equation}
Physically, $\etaref$ corresponds to the maximum displacement along the length of a {\em rigid} wing driven by $(\etaLE, \etaLE')$. This definition therefore applies to a wing driven by heaving, pitching, or a combination of the two.

\bibliographystyle{plain}
\bibliography{FlapBib}
\end{document}

%% file: Nomenclature.tex
\section{Nomenclature}
\label{NomenclatureSection}

\begin{table}[h]
\begin{center}
\label{symbols}
\begin{tabular}{| l l |}
\hline
Symbol 	& Definition \\
\hline

{\bf Dimensional quantities}& \\
$\amp, \freq, \Udim$	& the (peak-to-peak) flapping amplitude, frequency, and free-stream velocity \\
$\chord, \width, \thick$		& the wing's chord, span, and thickness respectively \\
$\mu(x), E(x), \Ia$		
& the wing's mass per unit length, elastic modulus, and second moment of area \\
$\rho, \rho_s$	& fluid and solid densities respectively (both mass per unit volume) \\
$\loadDim(x,t)$	& the dimensional hydrodynamic load \\

{\bf Dimensionless parameters}& \\
$\sigma$		& the reduced driving frequency, see Eq.~(\ref{ParamEq}) \\
$\UFS = {2 \pi}/{\sigma}$		& the dimensionless free-stream velocity \\
$\Mrat(x)$		& the ratio of solid-to-fluid inertia, see Eq.~(\ref{ParamEq}) \\
$\Stiff(x)$		& the dimensionless wing stiffness, see Eq.~(\ref{ParamEq}) \\
$\alpha(x)$, $\beta(x)$		
& variable coefficients that enter Eq.~(\ref{etaODE}); see Eq.~(\ref{alphabeta}) for definitions \\
$\etaLE$, $\etaLE'$		& the imposed heaving and pitching at the leading edge, see Eq.~(\ref{etaBC1}) \\ 
$\etalin(x) = \etaLE + \etaLE' (x+1)$	& a linear function, see Eq.~(\ref{etalin}) \\

{\bf Dimensionless variables}& \\
$h(x,t)$			& the wing's vertical displacement in Eq.~(\ref{beamND}) \\
$\eta(x)$			& the spatial component of $h(x,t)$, see Eq.~(\ref{hDecomp}) \\
$\eta_s(x)$			& the singular part of $\eta(x)$, see Eq.~(\ref{singODE0}) \\

$\load(x,t)$		& the dimensionless hydrodynamic load in Eq.~(\ref{beamND}) \\
$\loadx(x)$		& the spatial component of $\load$, see Eq.~(\ref{loadDecomp}) \\
$\loadx[\eta](x)$	& $\loadx$ considered as an operator on $\eta$, see Eq.~(\ref{etaODE})  \\
$\loadsing(x)$, $\loadreg(x)$	
& the singular and regular components of $\loadx(x)$, see Eq.~(\ref{loadDecomp2}) \\

$\paccel(x,y,t)$		& the Prandtl acceleration potential (i.e.~a normalized, negative pressure field) \\
$\psi(x,y,t)$		& the harmonic conjugate of $\paccel$ \\
$F(z,t) = \paccel + i \psi$	& the complex acceleration potential, see Eq.~(\ref{compvel}) \\
$w = u - iv$		& the complex velocity field, see Eq.~(\ref{compvel})  \\
$\Phi^{\pm}(x)$		& the spatial component of $\paccel$ evaluated on the wing surface \\
$\Psi(x)$			& the spatial component of $\psi$ evaluated on the wing surface \\
$V(x)$			& the spatial component of vertical velocity on the wing surface \\

{\bf Domains and operators}& \\
$\wing^{\pm} = \{ x \in [-1,1], y=0^{\pm} \}$		
& the top and bottom surfaces of the wing respectively \\
$ \ucirc = \{ \zeta = e^{i \theta} \}$	& the unit circle in the $\zeta$-plane \\
$\Dx = d / dx$		& differentiation with respect to $x$ \\
$\Popp [ \cdot ] = \Dx^2 \left( \alpha(x) \Dx^2 \cdot \right)$	
& the preconditioning operator, see Eq.~(\ref{Popp}) \\
$\Lopp$	& the linear operator defined in Eq.~(\ref{Lopp}) \\
{\bf Output quantities}& \\
$\CT$, $\CP$	& the thrust and power coefficients, see Eq.~(\ref{CTCPdefn}) \\
\hline
\end{tabular}
\end{center}
\end{table}